\documentclass[%
 reprint,
 amsmath,amssymb,
 aps,
]{revtex4-2}
\usepackage{subfigure}
\usepackage{color}
\usepackage{diagbox}
\usepackage{slashbox}
\usepackage{graphicx}
\usepackage{dcolumn}
\usepackage{bm}
\usepackage{natbib}
\usepackage{tabularx}

\begin{document}

\preprint{APS/123-QED}

\title{
Spectral separation of the stochastic gravitational-wave background for LISA: observing both cosmological and astrophysical backgrounds 
}%

\author{Guillaume Boileau}
 \email{guillaume.boileau@oca.eu}
\author{Nelson Christensen}%
 \email{nelson.christensen@oca.eu}
\affiliation{%
Artemis, Observatoire de la C\^{o}te d'Azur, Universit\'{e} C\^{o}te d'Azur, CNRS, CS 34229, F-06304 Nice Cedex 4, France
}%

\author{Renate Meyer}
 \affiliation{
 Department of Statistics, University of Auckland, Auckland, New Zealand} 
\email{renate.meyer@auckland.ac.nz}

\author{Neil J. Cornish}
\affiliation{%
 eXtreme Gravity Institute, Department of Physics,Montana State University, Bozeman, Montana 59717, USA
}%
\email{ncornish@montana.edu}

\date{\today}

\begin{abstract}
With the goal of observing a stochastic gravitational-wave-background (SGWB) with LISA, the spectral separability of the cosmological and astrophysical backgrounds is important to estimate. We attempt to determine the level with which a cosmological background can be observed given the predicted astrophysical background level. We predict detectable limits for the future LISA measurement of the SGWB. Adaptive Markov chain Monte-Carlo methods are used to produce estimates with the simulated data from the LISA Data challenge (LDC). We also calculate the Cramer-Rao lower bound on the variance of the SGWB parameter estimates based on the inverse Fisher Information using the Whittle likelihood. The estimation of the parameters is done with the 3 LISA channels $A$, $E$, and $T$. We simultaneously estimate the noise using a LISA noise model. Assuming the expected astrophysical background around $\Omega_{GW,astro}(25 \text{ Hz}) = 0.355 \rightarrow 35.5 \times 10^{-9}$, a cosmological SGWB normalized energy density of around $\Omega_{GW,Cosmo} \approx 1 \times 10^{-12}$ to $1 \times 10^{-13}$ can be detected by LISA after 4 years of observation.  
\begin{description}
 \item[keywords]
 \vspace{0.1cm}
Spectral separability, Stochastic gravitational-wave Background, LISA, Adaptive 
\newline Markov chains Monte-Carlo, Fisher Information, Whittle Likelihood.

\end{description}
\end{abstract}

\maketitle

\section{Introduction}

Since the accomplishment of the first detection of gravitational-waves from the merger of two stellar mass black
holes~\cite{Abbott:2016blz} by Advanced LIGO~\cite{Harry_2010,TheLIGOScientific:2014jea} and thereafter with Advanced Virgo~\cite{Acernese_2014,GW170814}, gravitational-wave observatories have become a new means to observe
astronomical phenomena. So far LIGO and Virgo have announced the observation of 50 signals produced from compact binary
coalescence~\cite{LIGOScientific:2018mvr,Abbott:2020niy}, including two from binary neutron star
mergers~\cite{TheLIGOScientific:2017qsa,Abbott:2020uma}. gravitational-wave detections are expanding our understanding
of astrophysics and of the Universe.  \\

The Laser Interferometer Space Antenna (LISA)~\cite{2017arXiv170200786A} is a future ESA mission, also supported by NASA, with the aim to observe gravitational-waves in the low-frequency band $[10^{-5}, 1]$ Hz. The mission lifetime will nominally be 4 years, but could be extendable to 6 or 10 years of scientific observations. LISA is a triangular constellation of three spacecraft, separated from one another at a distance of $L = 2.5 \times 10^9$ m.  
The low-frequency band is rich with gravitational-wave signals. The foreground of LISA will be dominated by sources from our galaxy, the Milky Way. White dwarf binaries \cite{10.1093/mnras/stz2834,PhysRevD.76.083006,Adams_2014} are numerous ($\sim$ 35 million binaries), and relatively near  the LISA constellation. For example, recently the Zwicky Transient Facility (ZTF) has measured a double white dwarf with an orbital period estimated at 7 minutes \cite{2019Natur.571..528B}, which corresponds to a gravitational-wave emission of $\simeq 30$ mHz. LISA can be expected to observe many resolved binaries, many of which are already known from photometry studies and constitute the so-called verification binaries \cite{2020ApJ...905...32B,Korol_2020}.  Well studied systems like this can be used to verify the LISA performance, acting as a way to confirm the sensitivity of LISA. We can expect to have one in a thousand binaries which are resolvable. The large majority of the galactic binaries are unresolved and form a stochastic signal. The stochastic gravitational-wave background from white dwarf binaries or galactic foreground will be anisotropic and the signal will be not a pure power law. 
A stochastic gravitational-wave background (SGWB)~\cite{Romano2017,Christensen_2018} will have a significant contribution from unresolved binaries, such as binary black holes and binary neutron stars. This background is essentially isotropic, and its level can be predicted from the signals observed by LIGO and Virgo \cite{2019ApJ...871...97C, PhysRevLett.116.131102}.  Another important SGWB would be from cosmological sources~\cite{Christensen_2018}. The origin of this background comes from the early Universe \cite{GarciaBellido:2007dg, 1995PhRvD..52.2083M}, with the possibility to measure  the inflation scenario parameters \cite{2020arXiv200704241C}. Cosmic strings could be another observable source~\cite{CHANG2020100604}. A cosmologically produced background can be modeled as a flat spectral energy density $\propto f^0$ \cite{Cornish_2001}. 

In this paper, we present a strategy to separate the two SGWBs (astrophysical and cosmological), as well as the LISA noise, using a Bayesian strategy \cite{PhysRevD.58.082001,PhysRevD.76.083006} based on an Adaptive Markov chain Monte-Carlo (A-MCMC) algorithm. We then show LISA's ability to measure a cosmological SGWB for different magnitudes for the astrophysical background. The SGWB from astrophysical sources today represents an important goal, especially considering the current observations by LIGO and Virgo~\cite{LIGOScientific:2019vic,Abbott:2021xxi}. 

Numerous studies have recently been presented which address how to possibly detect a cosmologically produced SGWB in the presence of an astrophysically produced SGWB. For example a recent study 
displays the use of principal component analysis to model and observe a SGWB in the presence of a foreground from binary black holes and binary neutron stars in the LISA observation band~\cite{Pieroni_2020}.
A component separation method is proposed in~\cite{2016JCAP...04..024P}, where they show that it is possible to detect an isotropic SGWB. 
The method uses maximum likelihood parameter estimation with Fisher Information matrices. This is proposed to replace an MCMC approach, and applied to the LIGO-Virgo observational band.

The proposal in~\cite{Ungarelli_2004} is to use a number of broken power-law filters to separate different backgrounds with gravitational-wave detectors on the Earth.
In the study of~\cite{Biscoveanu:2020gds} the proposal is to divide the data into individual short time segments. The method used the procedures described in~\cite{Smith:2017vfk} to search the segments for the presence of a binary black hole signal, either through direct detection or sub-threshold by generating a Bayesian evidence. A cosmological SGWB would be present in all segments, whereas a probability would exist for the presence of a binary black hole merger for the segments. The method is general, and could be applied to LIGO-Virgo or LISA. The study presented in~\cite{PhysRevLett.118.151105} noted that the sensitivity of third generation gravitational-wave detections, such as Einstein Telescope~\cite{Punturo_2010} or Cosmic Explorer~\cite{Reitze:2019iox}, will be so good that almost every binary black hole merger in the observable Universe can be directly detected, and then removed from the search for a cosmological SGWB. The study of~\cite{PhysRevD.102.063009} then explored how to do such a subtraction of binary black hole merger signals, and the consequences of the effect of residuals from such subtractions. Another study used Bayesian methods to address spectral separation for LIGO-Virgo observations, but trying to address how to separate a SGWB from a correlated magnetic noise background produced by the Schumann resonances~\cite{FULLEKRUG1995479,Sentman,Thrane:2013npa}; the study is, however, general and can be applied to spectral separation for different types of backgrounds~\cite{Meyers:2020qrb}. 
This study was then expanded to address the simultaneous estimation of astrophysical and cosmological SGWBs, and displayed that this will be especially important for third generation ground based detectors~\cite{Martinovic:2020}.
Another study, specifically dedicated to LISA observations~\cite{Caprini:2019pxz} proposes to divide the data into bins, and then within in each bin, a fit is made to a power law or a constant amplitude; a variation on this approach is presented here~\cite{1818908}. The claim is that this method is more dynamic and able to fit arbitrarly shaped SGWBs. The study of~\cite{10.1088/1361-6382/abb637} shows how to assign Bayes factors and probabilities to differentiate a SGWB signal from instrumental noise. 

All the SGWB studies referenced above are summarized in Tables \ref{table:introLIGO}, ~\ref{table:introLISA}, ~\ref{table:intro3gene}, respectively for LIGO/Virgo, LISA, and third generation detectors. We compare the goals, methods, the performance, the limitations and the application; see Appendix~\ref{sec:appendix}. The study we present in this paper, using Bayesian parameter estimation methods, has the advantage to fit two backgrounds and the LISA noise simultaneously. We note the possibility to expand the work presented here to estimate more complex LISA noise, and adding a new models for the SGWB; for example, more complex SGWBs could include broken power laws, peaks in the frequency domain, or an anisotropic SGWB from our galaxy.

The organization of the paper is as follows. In Sec.~\ref{spectral} we introduce the SGWB spectral separation problem for LISA, and then describe the inverse of the Fisher Information matrix of the SGWB parameters, and how this provides the Cramer-Rao lower bound on the variance of the parameter estimates. In Sec.~\ref{sc:adap_MCMC} we describe the A-MCMC. The simulated LISA mock data is presented in Sec.~\ref{sec:mock_data}. Presented in Sec.~\ref{sc:cov} are the parameter estimation procedures and results using the LISA $A$ and $T$ channels; Sec.~\ref{sec:AET} presents similar results using the LISA $A$, $E$ and $T$ channels.
Conclusions are given in Sec.~\ref{sec:conclusion}.

\section{Spectral Separation \label{spectral}}

An isotropic SGWB observed today $\Omega_{GW}(f)$ can be modeled with the frequency variation of the energy density of the gravitational-waves, $\rho_{GW}$, where $\mathrm{d} \rho_{GW}$ is the  gravitational-wave energy density contained in the frequency band $[f,\ f+df]$) \cite{doi:10.1146/annurev.nucl.54.070103.181251}. The distribution of the energy density over the frequency domain can be expressed as, 
\begin{equation}
\label{eq:Omega}
        \begin{split}
        \Omega_{GW}(f) &= \frac{f}{\rho_c} \frac{\mathrm{d} \rho_{GW}}{\mathrm{d} \ln(f)}\\
        &= \sum_k \Omega_{GW}^{(k)}(f)\\
        \end{split}
\end{equation}
where the critical density of the Universe is $\rho_c = \frac{3 H_0^2c^2}{8\pi G}$. In this paper we approximate the spectral energy density as a collection of power law contribution (this is a simplified model), $\Omega_{GW}(f) \simeq \sum_k A_{k}\left( \frac{f}{f_{ref}}\right)^{\alpha_k}$ where the energy spectral density amplitude of the component $k$ (representing the different SGWBs) is $A_k$, with the respective slope $\alpha_k$ and $f_{ref}$ is some characteristic frequency. The SGWB is predicted to have a slope component $\alpha \approx 0$ for the cosmological background. This is true for scale invariant processes, this is approximately true for the standard inflation and certainly false for cosmic string and turbulence. However for our study here we will model the cosmologically produced SGWB with $\alpha = 0$. In addition, we will use $\alpha = \frac{2}{3}$ for a compact binary produced astrophysical background. According to Farmer and Phinney the slope is $\alpha=\frac{2}{3}$ for quasi-circular binaries evolving purely under gravitational-wave emission~\cite{Farmer:2003pa}. The eccentricity and environmental effects can modify the slope. We also note the limitations of our power law model as phase transition in the early Universe can produce two-part power laws, with a traction between the rising and falling power law component at some peak frequency. But we start in this study with two power law backgrounds. As the two backgrounds are superimposed, the task is to  simultaneously extract both the astrophysical and cosmological properties, i.e.\ to simultaneously estimate the astrophysical and the cosmological contribution to the energy spectral density. 

To avoid identification issues, we use a Bayesian approach by putting informative priors on the individual slope and amplitude parameters.  Our work here builds on that of Adams and Cornish \cite{Adams_2010} where they demonstrated that it is possible to separate a SGWB from the instrumental noise in a Bayesian context. Similarly Adams and Cornish then showed that one could detect a cosmological SGWB in the presence of a background produced by white dwarf binaries in our galaxy \citep{Adams_2014}. Since the production of those studies LIGO and Virgo have observed gravitational-waves from binary black hole and binary neutron star coalescence. We now know that there will definitely be an astrophysically produced background across the LISA observation band produced by compact binary coalescences over the history of the Universe~\cite{2019ApJ...871...97C}, and if LISA is to observe a cosmologically produced background it will be necessary to separate the two. 

The literature displays large difference in the estimation of the magnitude of the astrophysically
produced SGWB. A recent simulation of the SGWB from merging compact binary sources with the \textbf{StarTrack} code \cite{2020arXiv200804890P} predicts an amplitude around  $\Omega_{GW} \simeq 4.97 \times 10^{-9}$ to $2.58 \times 10^{-8}$ at 25 Hz. However another study considered the binary black hole and binary neutron star observations by LIGO/Virgo, and produced predictions going from the LISA observational band to the LIGO/Virgo band. They estimate an amplitude for the astrophysical SGWB of  $\Omega_{GW} \simeq 1.8 \times 10^{-9}$ to $2.5 \times 10^{-9}$ at 25 Hz \cite{2019ApJ...871...97C}. These amplitudes can be propagated to the LISA band by recalling Eq.~\ref{eq:Omega} and using $f_{ref} = 25$ Hz and $\alpha = 2/3$. In the context of an effort to observe a cosmological SGWB we have large variations due to the predictions of the astrophysical component. 

In our study here we predict the accuracy of a measurement of $\Omega_{GW}^{(0)}$ with astrophysical inputs of differing magnitudes using $f_{ref} = 25$ Hz, $\Omega_{GW}^{(\frac{2}{3})} = $ $[3.55 \times 10^{-10},$  $1.8 \times 10^{-9},$  $3.55 \times 10^{-9},$  $3.55 \times 10^{-8}]$ after 4 years of observation. 
We use the  orthogonal LISA  $A$, $E$, and $T$ channels, which are created from the time delay interferometry (TDI) variables $X$, $Y$, and $Z$~\cite{PhysRevD.66.122002}.
Our method fits the parameters of two stochastic backgrounds, and simultaneously the LISA noise with the help of the channel $T$. We assume uncorrelated noise TDIs between the "science" channels ($A,E$) and the noise channel ($T$). The $T$ channel is "signal insensitive" for gravitational-wave wavelengths larger than the arm lengths. The noise channel $T$ is obtained from a linear combination~\cite{PhysRevD.66.122002} of the TDIs channel $(X,Y,Z)$. We demonstrate a good ability to estimate the noise present in the two science data channels $A$ and $E$. We can then set a limit on the ability to detect the cosmological SGWB. The predictions from the Bayesian study are confirmed via a study of the frequentist estimation of the error. Namely, we use a Fisher information analysis, performed for the spectral separation independently of the Bayesian A-MCMC approach. The inverse of the Fisher Information matrix of the SGWB parameters, presented in Sec.~\ref{spectral}, provides the Cramer-Rao lower bound on the variance of the SGWB parameter estimates.

A useful toy model to consider is the problem of separating two independent stationary mean-zero Gaussian noise processes that have different power spectra $S_{n_1} (f) = A_1f^{\alpha_1}$ and $S_{n_2}(f) = A_2f^{\alpha_2}$. Suppose we have data that is formed from the sum of these two independent noise processes 
\begin{equation}
d(t)=n_1(t) +n_2(t),\quad t=1,\ldots,T.
\end{equation}
After a Fourier transform to $\tilde{d}(f_k)=\frac{1}{\sqrt{T}}\sum_{i=1}^T d(t) e^{-itf_k}$ at Fourier frequencies $f_k= 2\pi k/T,\; k=0,\ldots, N=\frac{T}{2}-1$ (for $T$ even), 
 we can write:

\begin{equation}
    \tilde{d}(f_k) = \tilde{n}_1(f_k) +\tilde{n}_2(f_k),\quad k=0,\ldots,N.
\end{equation}
Then the vector $\tilde{d}$ has an asymptotic complex multivariate Gaussian
distribution with a diagonal covariance matrix. The diagonal elements are given by the values of the spectral density $S(f_k)= A_1f_k^{\alpha_1} + A_2f_k^{\alpha_2}$. Our assumption of independence implies that one can simply sum the individual spectral densities of the two noise processes.

The Whittle likelihood approximation in the frequency domain can then be written as:
\begin{equation} \label{likelihood}
    p(d|A_1 , \alpha_1 , A_2 , \alpha_2) = \prod_{k=1}^N \frac{1}{\pi S(f_k)} e^{-\frac{\tilde{d}(f_k)^{\star} \tilde{d}(f_k)}{ S(f_k)}}
\end{equation}
where $S(f_k ) = A_1f_k^{\alpha_1} + A_2f_k^{\alpha_2}$. 
The product $I_n(f_k)=\tilde{d}(f_k)^{\star} \tilde{d}(f_k)$ is the {\em periodogram}, the squared magnitude  of the Fourier coefficients at the frequency $f_k$.
The log likelihood (up to an additive constant) is thus
\begin{equation}
    \ln p(d|A_1 , \alpha_1 , A_2 , \alpha_2) = - \sum_{k=1}^N \left( \frac{I_n(f_k)}{S(f_k)} + \ln S(f_k) \right) .
\end{equation}

\subsection{The Fisher information}\label{sc:FisherInfo}
The Fisher information matrix $\Gamma$ for a parameter vector $\boldmath{\theta}=(\theta_1,\ldots,\theta_p)$ is given by the expected value of the negative Hessian of the
log likelihood. The element in row $i$ and column $j$ of the Fisher information is given by:
\begin{equation}
\Gamma_{ij}=E\left[ -\frac{\partial^2}{\partial \theta_i \partial \theta_j } \ln p(d|\boldmath{\theta})\right]
\end{equation}
The Fisher information can be easily obtained for the parameter vector
$(A_1,\alpha_1,A_2,\alpha_2)$ by using that (asymptotically) $E[I_n(f_k)]=S(f_k)$ and $\Gamma_{ij}=\Gamma_{ji}$. 
\begin{eqnarray}\label{Fisher}
\Gamma_{11}&=& 
 \sum_{k=1}^N \frac{f_k^{2\alpha_1}}{(A_1f_k^{\alpha_1} + A_2 f_k^{\alpha_2})^2}\\
  \Gamma_{22} &=& \sum_{k=1}^N \frac{ (A_1f_k^{ \alpha_1} \ln f_k )^2}{(A_1f_k^{ \alpha_1} +A_2f_k^{ \alpha_2})^2} \\
  \Gamma_{33}&=& 
 \sum_{k=1}^N \frac{f_k^{2\alpha_2}}{(A_1f_k^{\alpha_1} + A_2 f_k^{\alpha_2})^2}\\
  \Gamma_{44} &=& \sum_{k=1}^N \frac{ (A_2f_k^{ \alpha_2} \ln f_k )^2}{(A_1f_k^{ \alpha_1} +A_2f_k^{ \alpha_2})^2} \\
\Gamma_{12}&=&\Gamma_{21}= \sum_{k=1}^N \frac{A_1f_k^{2\alpha_1} \ln f_k }{(A_1f_k^{ \alpha_1} +A_2f_k^{ \alpha_2})^2} \\
\Gamma_{13}&=&\Gamma_{31}= \sum_{k=1}^N \frac{f_k^{\alpha_1+\alpha_2}}{(A_1f_k^{ \alpha_1} +A_2f_k^{ \alpha_2})^2} \\
\Gamma_{14}&=&\Gamma_{41}= \sum_{k=1}^N \frac{A_2f_k^{\alpha_1+\alpha_2} \ln f_k }{(A_1f_k^{ \alpha_1} +A_2f_k^{ \alpha_2})^2} \\
\Gamma_{23}&=&\Gamma_{32}= \sum_{k=1}^N \frac{A_1f_k^{\alpha_1+\alpha_2} \ln f_k }{(A_1f_k^{ \alpha_1} +A_2f_k^{ \alpha_2})^2} \\
\Gamma_{24}&=&\Gamma_{42}= \sum_{k=1}^N A_1A_2\frac{A_1A_2f_k^{\alpha_1+\alpha_2}\ln^2 f_k}{(A_1f_k^{ \alpha_1} +A_2f_k^{ \alpha_2})^2} \\
\Gamma_{34}&=&\Gamma_{43}= \sum_{k=1}^N \frac{A_2f_k^{2\alpha_2} \ln f_k }{(A_1f_k^{ \alpha_1} +A_2f_k^{ \alpha_2})^2} 
\end{eqnarray}

\subsection{The Cramer-Rao bound}\label{sec:Cramer}
The Fisher information can be used to give a lower bound for the variance of any unbiased estimator, the so called Cramer-Rao bound.
For  any unbiased estimator $ \widehat{\theta_i}$ of the unknown parameter $\theta_i$, its standard error $\Delta \widehat{\theta}_i $ satisfies
\begin{equation}
    (\Delta \widehat{\theta_i})^2 \ge \Gamma_{ii}(\theta)^{-1} = \frac{1}{E\left[-\frac{\partial}{\partial \theta_i}\frac{\partial}{\partial \theta_i} \ln{p(d|\theta)}\right]}
\end{equation}
Under certain regularity conditions, the posterior distribution of a parameter $\theta$ is asymptotically Gaussian, centered at the posterior mode and
covariance matrix equal to the inverse of the negative Hessian of the posterior distribution evaluated at the posterior mode. For flat priors, the posterior density is proportional to the likelihood, the posterior mode is the maximum likelihood estimate and the 
 standard error $\Delta \widehat{\theta}_i $ of the Bayesian estimator  $\widehat{\theta}_i$ of the parameter $\theta_i$ can  be approximated by evaluating the  Fisher information at $\widehat{\theta}_i$, i.e.
\begin{equation}
  \Delta \widehat{\theta}_i  \approx \Gamma_{ii}(\widehat{\theta}_i)^{-1/2}.
\end{equation}
Defining the {\em uncertainty} of an estimate $\widehat{\theta_i}$ by 
\begin{equation}
    \frac{\Delta \widehat{\theta}_i}{\widehat{\theta}_i}  
\end{equation}
we say that we can estimate the parameter $\theta_i$ with on error of $10\%$ based on the Fisher analysis if the uncertainty of a parameter estimate is equal to $0.1$.
 The purpose of this study is to derive a  threshold on the separability by an A-MCMC routine with the likelihood of the Eq.~\ref{likelihood}. In the following we will thus have a limiting value for the separability of the cosmological SGWB parameters and the astrophysical SGWB.

\begin{figure*}[t!]
    \centering
    \includegraphics[height= 9cm]{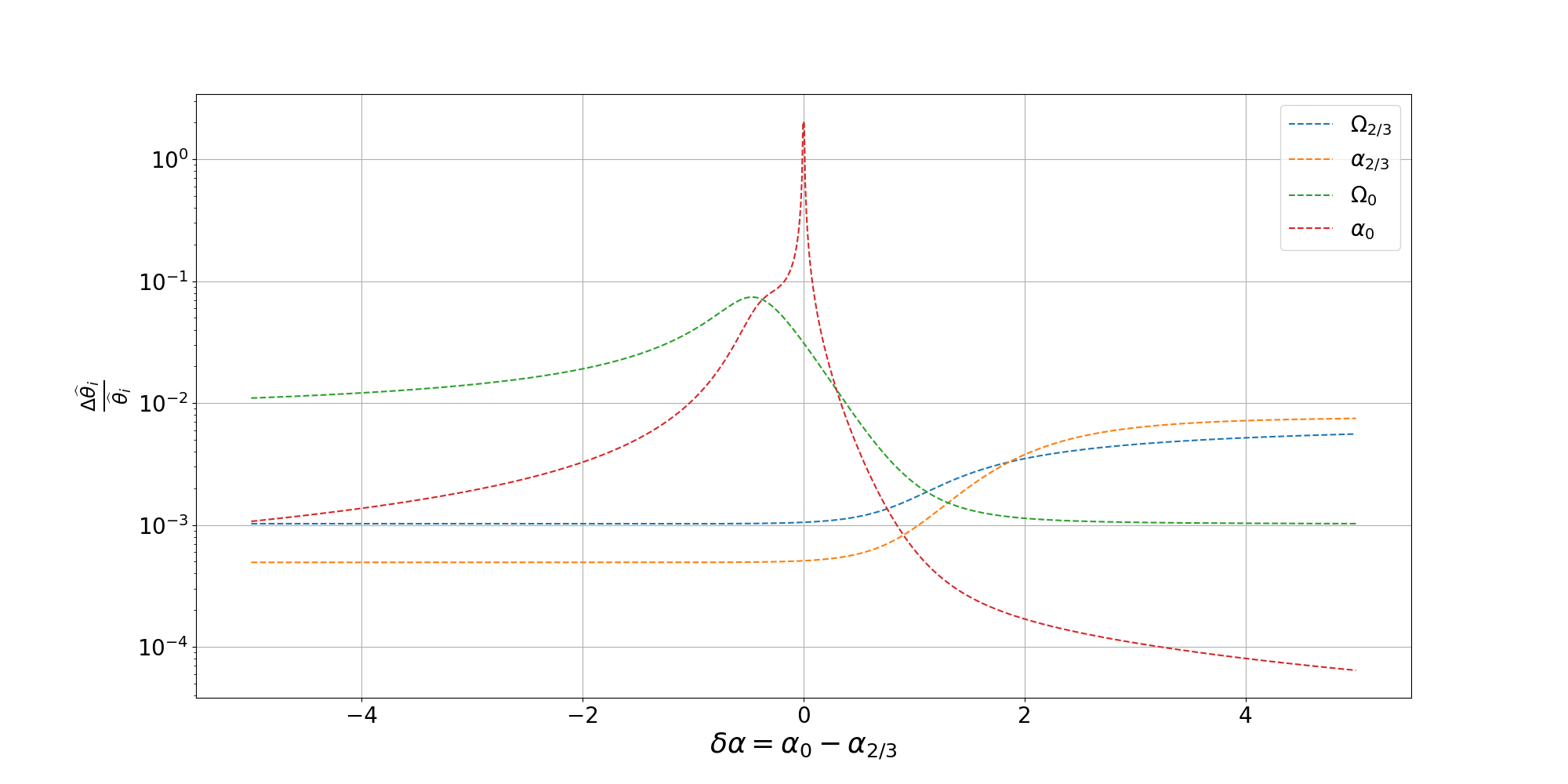}
    \caption{Uncertainties ($\frac{\Delta \widehat{\theta}_i}{\widehat{\theta}_i}$) of the amplitudes and spectral slopes as a function of the difference in the differential spectral slopes ($\delta \alpha = \alpha_0 - \alpha_{2/3}$).}
    \label{fig:deltaalpha}
\end{figure*}

We use a toy problem to display the separability of two stochastic backgrounds according to their slope difference. For this we fix one background $\Omega_1(f) = A_1 \left(\frac{f}{f_{ref}}\right)^{\alpha_1} = \Omega_{2/3} \left(\frac{f}{f_{ref}}\right)^{\alpha_{2/3}} = 3.55 \times 10^{-9} \left(\frac{f}{25 \text{Hz}}\right)^{2/3}$, and we leave free the slope of the second background $\Omega_2(f) = A_2 \left(\frac{f}{f_{ref}}\right)^{\alpha_2} =\Omega_0 \left(\frac{f}{f_{ref}}\right)^{\alpha_0} = 1 \times 10^{-12} \left(\frac{f}{25 \text{Hz}}\right)^{\alpha_0}$. We show the uncertainties ($\frac{\Delta \widehat{\theta}_i}{\widehat{\theta}_i}$ for $\theta_i \in [\Omega_{2/3}, \alpha_{2/3}, \Omega_0, \alpha_0]$, with $\Delta \widehat{\theta}_i$ the error from the Fisher information, see Sec.~\ref{sec:Cramer}) for the amplitudes and spectral slopes as a function of the difference between the spectral slopes ($\delta \alpha = \alpha_0 - \alpha_{2/3}$). This quantity is also called coefficient of variation or the relative standard deviation (RSE), this is the absolute value of the standard deviation divided by the mean of the parameter. We use this quantity to appreciate the dispersion of values around the mean. it is preferable to use this quantity because it is unitless. Thus it is easier to compare parameters of different units and ranges values.  Fig.~\ref{fig:deltaalpha} displays the uncertainties ($ \frac{\Delta \widehat{\theta}_i}{\widehat{\theta}_i}$) as the function of $\delta \alpha$ between -5 and 5.   

The uncertainty of the parameter $\alpha_0$ becomes larger when the slope difference $\delta \alpha$ is near to zero. Here it is more difficult to separate the two backgrounds when their slopes are similar. The uncertainties are also not symmetric about $\delta \alpha = 0$ because when the slope changes the amplitude is also changing by a factor $f_{ref}^{-\alpha}$. The uncertainty of the amplitude parameter $\Omega_0$ is maximal when the two amplitude parameters are identical. The position of the maximum changes for different inputs of $\Omega_0$; if $\Omega_0$ increases the position of the maximum converge to $\delta \alpha = 0$.

\section{Adaptive Markov chain Monte-Carlo} \label{sc:adap_MCMC}
\subsection{Markov chain Monte-Carlo}
Bayesian inference quantifies the estimation and uncertainties of unknown parameters 
 based on the observation of events that depend on these parameters. The quantification uses the {\em posterior} probability distribution.
It is obtained using Bayes' theorem (see Eq.~ \ref{eq:bayes}) by updating the {\em prior} distribution of the parameters with the {\em likelihood} $p(d|\theta)$, the conditional distribution of the observations given the parameters:
\begin{equation}\label{eq:bayes}
    p(\theta|d) = \frac{p(d|\theta)p(\theta)}{p(d)}
\end{equation}
where $p(\theta)$ is the prior distribution, $p(\theta|d)$ is the posterior distribution, and $p(d)=\int p(d|\theta)p(\theta) d\theta$ is the {\em evidence}. 

MCMC methods \cite{GelmanAndrew2014} provide a numerical strategy to compute the joint posterior distribution and its marginal distributions. It is a sampling-based approach that simulates a Markov chain constructed in such a way that its invariant distribution is the joint posterior.

\subsection{Metropolis-Hasting sampler}
As it is generally difficult to sample independently from a multivariate distribution,
MCMC methods draw dependent samples from Markov chains. 
The predominant MCMC algorithm is the 
Metropolis-Hastings (MH) algorithm. It is based on the rejection or acceptance of a candidate parameter $\theta'$ where the acceptance probability is given by  likelihood ratio
between the candidate and the previously sampled parameter value. 
Thus, any move into the direction of higher likelihood (towards the maximum likelihood estimation) will always be accepted, but because downhill moves still have a chance to be accepted, the MH algorithm avoids getting stuck in local maxima. \newline

\textbf{Metropolis-Hastings algorithm}
    \begin{itemize}
        \item Randomly select an initial point $\theta^{(0)}$.
        \item At the $n$th iteration:
            \begin{itemize}
                \item Generation of candidate $\theta'$ with the proposal distribution $g(\theta'|\theta^{(n)})$ 
                \item Calculation of acceptance probability\\
                    $\alpha = \min \left[1,\frac{p(d|\theta')}{p(d|\theta^{(n)})} \frac{p(\theta{(n)})}{p(\theta')}\right]$
                  
                \item Accept/Reject
                \begin{itemize}
                    \item Generation of a uniform random number $u$ on $[0,1]$
                    \item if $u \leq \alpha$, accept the candidate:\\ $\theta^{(n+1)} = \theta'$
                    \item if $u > \alpha$, reject the candidate:\\ $\theta^{(n+1)} = \theta^{(n)}$
                \end{itemize}
            \end{itemize}
    \end{itemize}
    Note that the proposal distribution $g$ is often chosen to be Gaussian centered around the current parameter value.
While executing the algorithm, we can monitor the acceptance rate, the proportion of candidates that were accepted. On the one hand, if this number is too close to 0 then the algorithm makes large moves into the tails of the posterior distribution which have low acceptance probability causing the chain to stay at one value for a long time. On the other hand,  a high acceptance rate indicates that the chain makes only small moves causing slow mixing. To control the mixing of the Markov chain we can introduce an adaptive step-size parameter that controls the size of the moves; this is the standard deviation in case of a univariate Gaussian proposal or the covariance matrix of a multivariate Gaussian proposal. As the iterations of the algorithm proceed, it is possible to dynamically modify the step-size to improve the  convergence of the chain. Intuitively, an optimal proposal would be as close to the posterior distribution as possible. Using a Gaussian proposal, its covariance matrix should thus be as close to the covariance matrix of the posterior distribution. Since the previous MCMC samples can be used to provide a consistent estimate of the covariance matrix, this estimate can be used to adapt the proposal on the fly, as detailed in \ref{sec:adaptive}.

\subsection{Adaptive Markov chain Monte-Carlo} \label{sec:adaptive}

We use the version of the Adaptive Metropolis MCMC from Robert and Rosenthal~\cite{10.1198/jcgs.2009.06134}. 
For a $p$-dimensional MCMC we can perform the Metropolis-Hasting with a  proposal density $g_n(.|\theta^{(n)})$ in iteration $n$ defined by a mixture of Gaussian proposals:
\begin{equation}
\begin{split}
    g_n(.|\theta^{(n)}) =& (1-\beta)\,N\Bigg(\theta^{(n)},\frac{(2.28)^2}{p} \Sigma_n \Bigg) \\
    &+ \beta \, N\Bigg(\theta^{(n)},\frac{(0.1)^2}{p} I_p \Bigg)
\end{split}
\end{equation}
with $\Sigma_n$ the current empirical estimate of the covariance matrix, $\beta = 0.25$ a constant, $p$ the dimensionality of the parameter space, $N$ the multi-normal distribution and $I_p$ the $p\times p$ identity matrix. We compute an estimate $\Sigma_n$ of the covariance matrix using the last hundred samples of the chain. The chain generated from an adaptive algorithm is not Markovian but the diminishing adaptation condition ensures ergodicity and thus the convergence to the stationary distribution.

\section{Data from the Mock LISA Data Challenge}
\label{sec:mock_data}

\subsection{Noise and SGWB energy spectral density of the MLDC}
\begin{figure*}[t]
    \centering
    \includegraphics[height= 9cm]{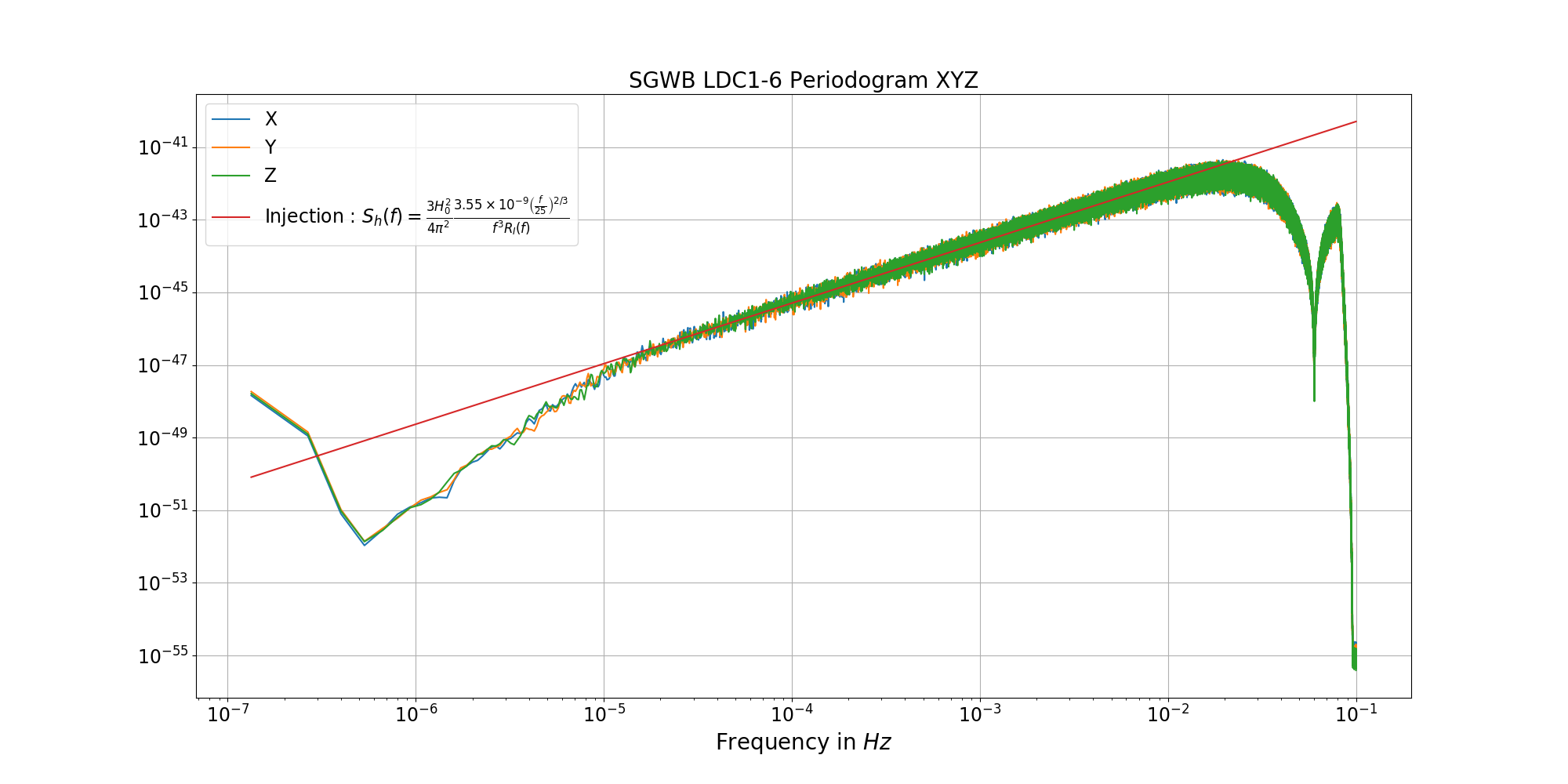}
    \caption{Periodogram of the channels $(X,Y,Z)$ of the SGWB from MLDC (\textit{LDC1-6} noiseless) with a single background ($\Omega_{GW}(f) = 3.55 \times 10^{-9} \left( \frac{f}{25 \text{Hz}}\right)^{2/3}$)}
    \label{fig:T-SXYZ}
\end{figure*}

\begin{figure*}[t]
    \centering
    \includegraphics[height= 9cm]{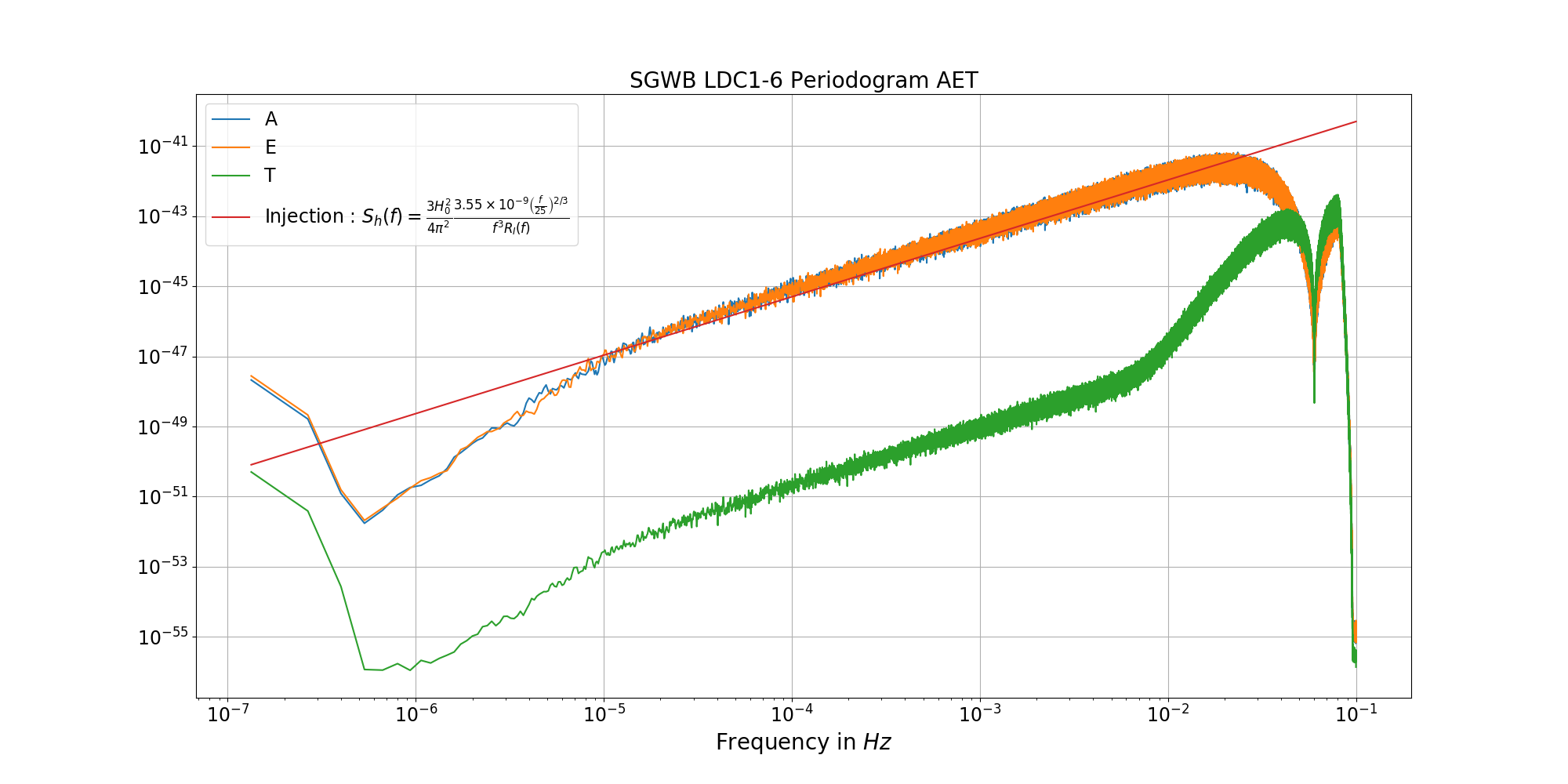}
    \caption{Periodogram of the channels $(A,E,T)$ of the SGWB from MLDC (\textit{LDC1-6} noiseless) with an single background ($\Omega_{GW}(f) = 3.55 \times 10^{-9} \left( \frac{f}{25 \text{Hz}}\right)^{2/3}$)}
    \label{fig:T-SAET}
\end{figure*}

The mock LISA data challenge (MLDC) provides simulations of  the  signal and noise of LISA in the approximation of one arm. We use the $(X,Y,Z)$ time series of the \textit{LDC1-6} data set from the MLDC webpage \cite{LDCM}. These are simulations of a binary produced SGWB of the form
$\Omega_{GW}(f) = \Omega_{2/3} \left(\frac{f}{f_{ref}} \right)^{\alpha}$ for $f_{ref} = 25$ Hz with a slope $\alpha = \frac{2}{3}$ and an amplitude of $\Omega_{2/3} = 3.55 \times 10^{-9} \ (\text{at } 25 \ \text{Hz)})$. Fig.~\ref{fig:T-SXYZ} and \ref{fig:T-SAET} display the gravitational-wave periodograms for the $(X,Y,Z)$ and $(A,E,T)$ channels.

We can transform the $X,Y,Z$ time series to the $A,E,T$ channels according to:
\begin{equation} 
    \left\{
\begin{array}{l}
     A = \frac{1}{\sqrt{2}}(Z-X) \\
     E = \frac{1}{\sqrt{6}}(X-2Y+Z) \\
     T = \frac{1}{\sqrt{3}}(X+Y+Z).
\end{array}
\right.
\label{eq:AET} 
\end{equation}
This linear combination of the original channels used to define $T$ has been shown to be insensitive to the gravitational-wave signal. While this is not exactly true, we will maintain that assumption for this analysis. As such, $T$ can be regarded as a null channel which contains mainly only noise, while channels $A$ and $E$ are the science channels, containing the gravitational-wave signal in the presence of noise~\cite{Romano2017}. In the following we focus on the science channels, $A$ and $E$.

In this study we use a simplified model where we assume equal noise levels on each spacecraft. According to Adams and Cornish \cite{Adams_2014} one can use a more complicated model that allowed for different noise levels. Future work will address this, plus the situation where the slope parameters for the noise can also vary. These parameters could then also be estimated by Bayesian parameter estimation methods.  

\begin{figure*}[t]
    \centering
    \includegraphics[height= 6cm]{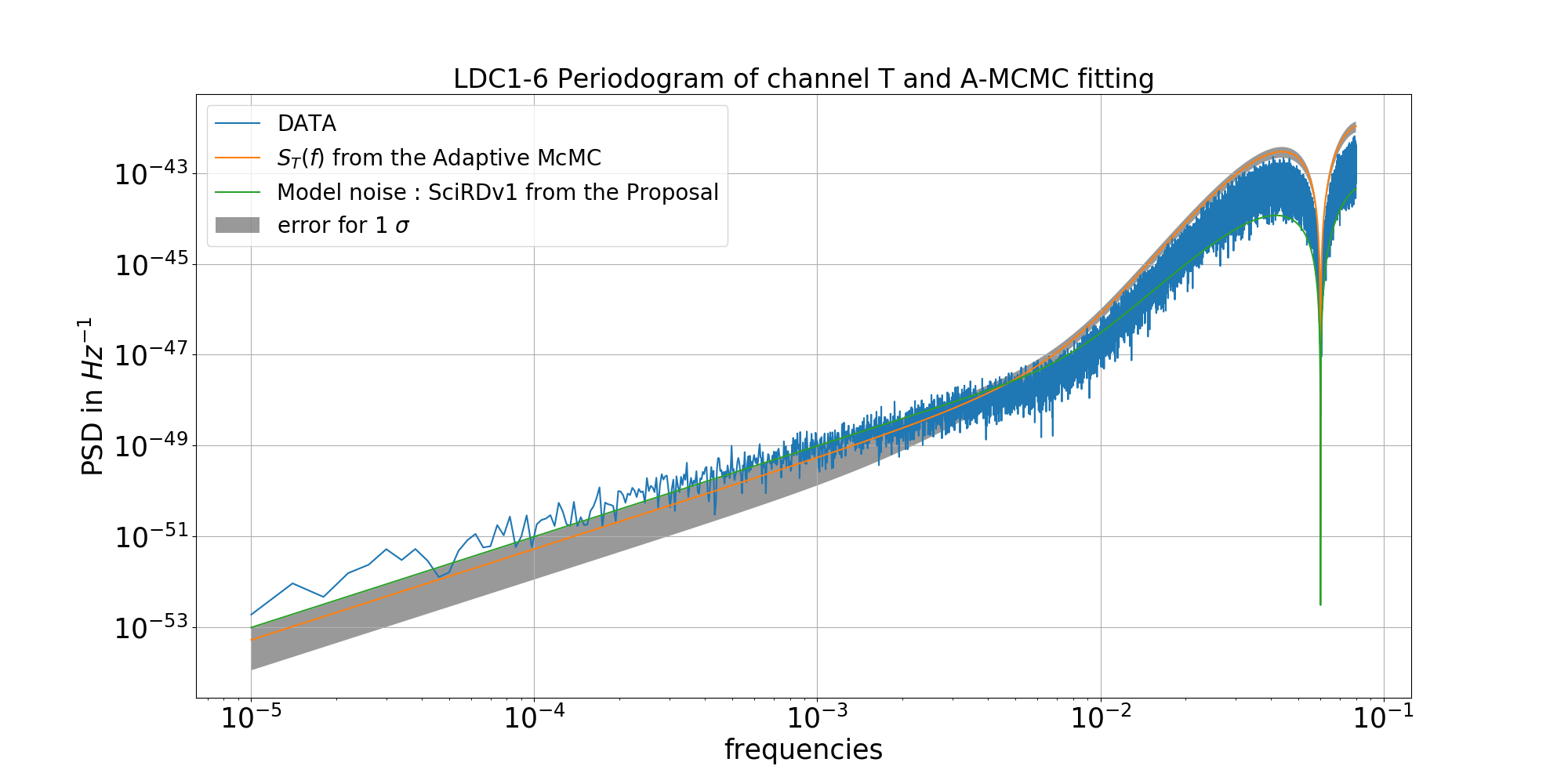}
    \caption{Power spectral density of the channel $T$ from the MLDC (in blue)~\cite{LDCM}. The green line represents the analytic noise model of the power spectral density of the channel $T$ with the parameters from the proposal~\cite{LSR}. 
    The orange line is the model from Eq.~\ref{eq:modelPSDT} with the values fit with the MCMC. In grey is the 1 $\sigma$ error. This is the uncertainty calculated from Eq.~\ref{eq:errorPSD}, where we take $dPSD_T$ with $dN_{pos} = \sigma_{N_{pos}}$ and $dN_{acc} = \sigma_{N_{acc}}$; $\sigma$ is the standard deviation of the posterior estimation. See Fig.~\ref{fig:corner_PSDT} and Eq.~\ref{eq:PSDTresult}.} 
    \label{fig:MCMC_PSDT}
\end{figure*}

\begin{figure}[b]
    \centering
    \includegraphics[height= 7cm]{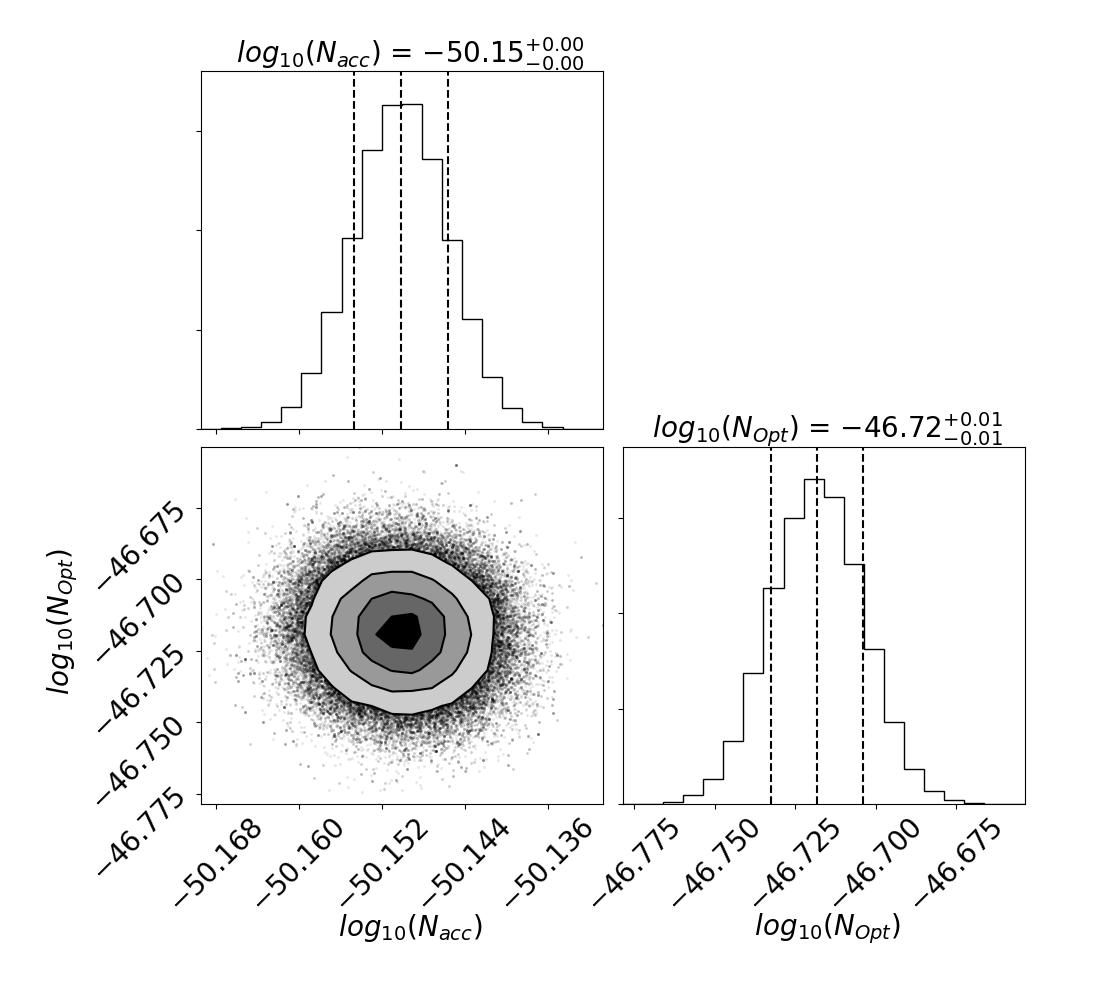}
    \caption{Corner plot for the A-MCMC generated posterior distributions for the power spectral density of the channel $T$ of the MLDC data set, estimating the two magnitudes of the LISA noise model from the proposal ~\cite{LSR}. The vertical dashed lines on the posterior distributions represent, from left to right, the quantiles [16\%, 50\%, 84\%]. }
    \label{fig:corner_PSDT}
\end{figure}

For the following studies we restrict the frequency band to correspond to the LISA band $[10^{-5},1]$ Hz. The power spectral density (PSD) of the channel $T$, $S_T$, can be described as (according to~\cite{LDCM}):
\begin{equation} \label{eq:modelPSDT}
    \begin{split}
        S_T(x) =& 16 S_{Op}(x)\left(1 - \cos(x)\right) \sin^2 (x) \\ 
        &+128 S_{pm}(x) \sin^2 (x) \sin^4\left(\frac{x}{2}\right)
    \end{split}
\end{equation}
with $x = \frac{2\pi L}{c}f$, $S_{Op}$ is the optical metrology system noise and $S_{pm}$ is the acceleration and displacement noise. The LISA noise budget is: 
\begin{equation}
\left\{
\begin{array}{l}
    S_{Op}(f) = N_{Opt}L^2\left(1 + \left(\frac{8 \text{ mHz}}{f} \right)^4 \right) \\
    S_{Pm}(f) = N_{Acc} L^2 S_{Acc}(f) S_{Dis}(f)
\end{array}
\right.
\end{equation}
with 
\begin{equation}
\left\{
\begin{array}{l}
    S_{Acc}(f) = \left(1 + \left(\frac{0.4 \text{ mHz}}{f} \right)^2 \right) \left(1 + \frac{f}{8 \text{ mHz}}\right)^4 \\
    S_{Dis}(f) = \left( 2 \pi f\right)^{-4}\left( \frac{2 \pi f}{c} \right)^2
\end{array}
\right.
\end{equation}    
The two free parameters, $N_{Opt}$ and $N_{Acc}$, are the respective levels of the two principal sources of noise in the LISA noise budget. In the LISA Science Requirement Document~\cite{LSR}, the level of the LISA noise acceleration is $N_{Acc} = 1.44 \times 10^{-48} \ \text{s}^{-4} \text{Hz}^{-1}$ and the upper limit on the level of the optical metrology system noise is $N_{Opt} = 3.6 \times 10^{-47} \ \text{Hz}^{-1}$. From the modeling of the strain requirements of the mission performance requirements, this is a maximisation of the noise level. The LISA noise budget corresponds to all sources of contamination that contribute to the power spectral density of the LISA detection system. The two noise sources correspond to estimates of different physical effects. We clearly do not yet have the true values for these physical effects; we presently only have estimates from experiments. The LISA requirements fixed the limit of the two magnitude levels so as to respect LISA's detection performance.
In Fig.~\ref{fig:MCMC_PSDT}, the green curve is the analytic noise model of the PSD of the channel $T$ with the parameters from the proposal~\cite{LSR}.The blue curve is the periodogram for the channel $T$ of the MLDC data ($LDC1-6$ SGWB signal); this is the magnitude squared of the Fourier coefficients for the (see Eq.~\ref{eq:AET}) data. Assuming the functional form of the noise PSD in channel $T$ is given by (\ref{eq:modelPSDT}), we can use the A-MCMC (see Sec.~\ref{sc:adap_MCMC}) to  fit the LISA Noise Parameters $N_{Opt}$ and $N_{Acc}$.
 The priors for the two components are flat log-uniform distributions and we specify $\beta = 0.01$ and $N = 200 000$ in the A-MCMC algorithm.  
The orange curve in Fig.~\ref{fig:MCMC_PSDT} is the estimated PSD based  on  Eq~\ref{eq:modelPSDT} with $N_{Opt}$ and $N_{Acc}$ replaced by the posterior means of samples obtained via the A-MCMC, given in Eq~\ref{eq:PSDTresult}. The 1 $\sigma$ error bands are overlaid in grey. 
Fig.~\ref{fig:corner_PSDT} shows the corner plot for the posterior samples of the two parameters, and the empirical posterior distributions seem to be well approximated by Gaussian distributions. 
It shows that this model yields a reasonable fit to the simulated channel $T$ data. 
We acknowledge that this is a rigid noise model for the purpose of this study, and future work will include more realistic scenarios: allowing for different noise levels on each spacecraft~\cite{Adams_2014}, allowing for small modifications of the transfer functions, and  allowing for small modifications in the spectral slopes of the noise components.
The posterior means of the two noise parameters are:

\begin{equation}\label{eq:PSDTresult}
\left\{
\begin{array}{l}
    \widehat{N}_{acc} = 7.08 \times 10^{-51} \pm 4 \times 10^{-53} \ \text{s}^{-4} \text{Hz}^{-1} \\
   \widehat{N}_{Opt} = 1.91 \times 10^{-47} \pm 4 \times 10^{-49} \ \text{Hz}^{-1}
\end{array}
\right.
\end{equation}
The gravitational-wave energy spectral density $\Omega_{GW}$ can be defined as 
\begin{equation}
	\Omega_{GW,I}(f) = \frac{2 \pi^2 }{3 H_0^2}f^3 \frac{PSD_I(f)}{R_I(f)}
\end{equation}
for $I=A, E$, where $H_0$ the Hubble-Lemaître constant ($H_0 \simeq 2.175 \times 10^{-18}  \ \text{Hz}$), $PSD_I$ the power spectral density of the channel $I$ and $R_I$ the response function. An asymptotically unbiased estimate of $PSD_I$ is given by the periodogram $I_n(f)=\sum_{k=1}^{N} |\tilde{d}(f_k)|^2 = \tilde{d}_I^*(f_k) \tilde{d}_I(f_k)$.

\begin{figure*}
    \centering
    \subfigure[Total frequency band of Channels A and E]{\includegraphics[width=0.47\textwidth]{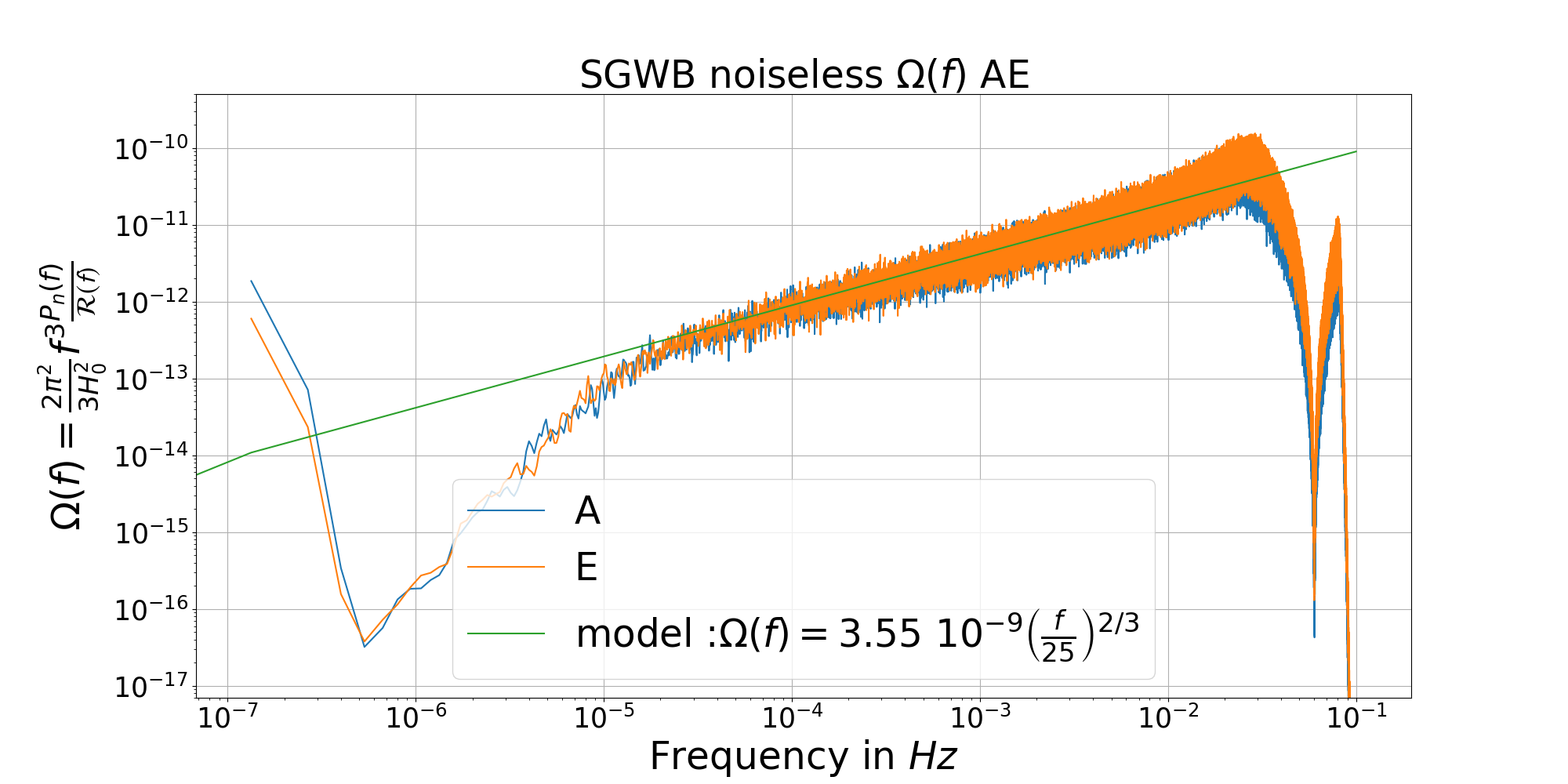}}
    \subfigure[Reduce frequency band $2.15 \times 10^{-5}$ to $9.98 \times 10^{-3} \ \text{Hz}$ of Channels A and E]{\includegraphics[width=0.47\textwidth]{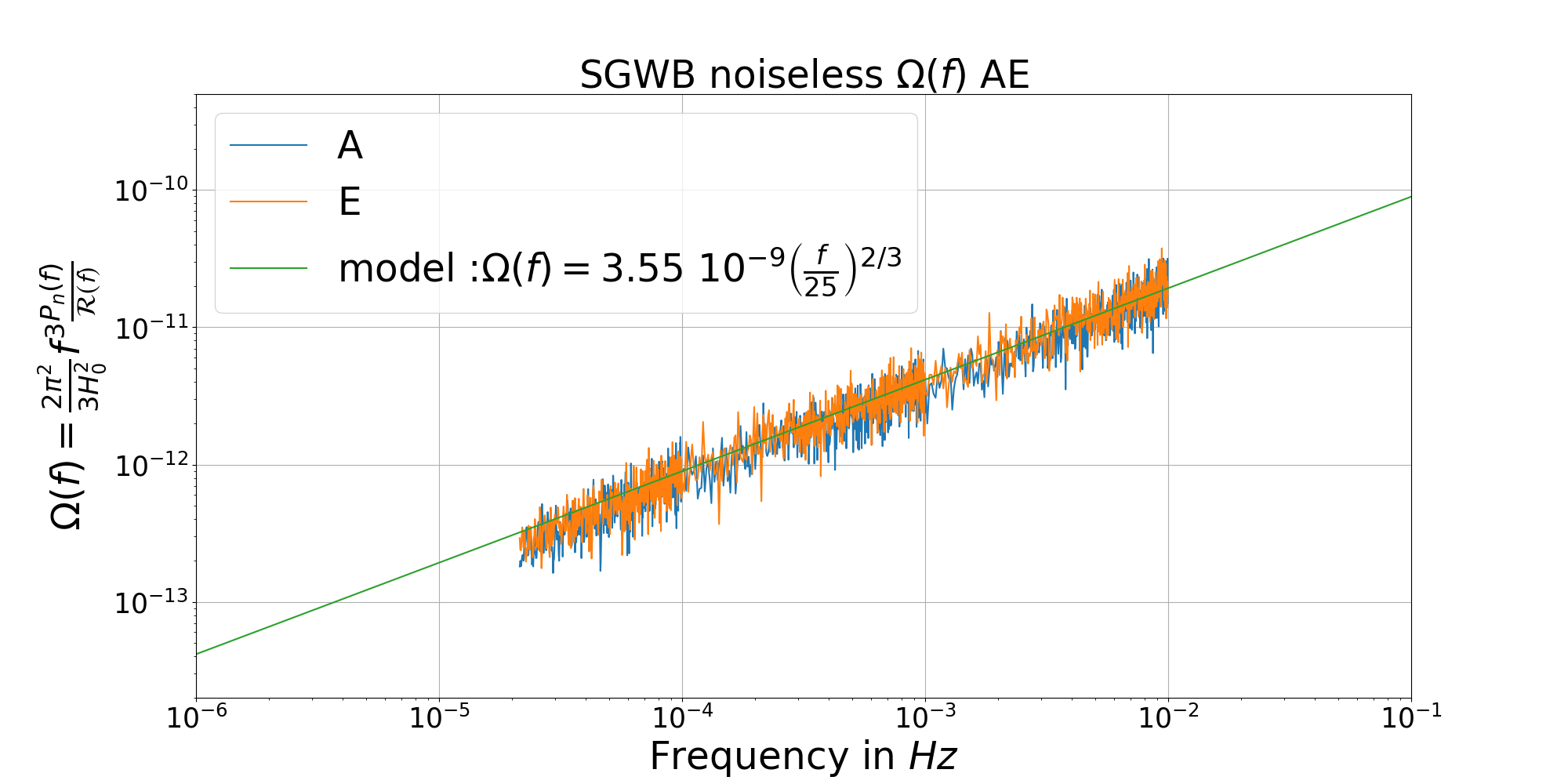}} 
    \caption{ Observations in channels [$A$,$E$] of the spectral energy density of the SGWB from astrophysical background $\Omega_{GW}(f)$ of the MLDC for the noiseless channel, Eq.~\ref{eq:R_noiseless}. (a) Total frequency band of Channels $A$ and $E$. (b) Reduced frequency band $2.15 \times 10^{-5}$ to $9.98 \times 10^{-3} \ \text{Hz}$ of Channels $A$ and $E$.}\label{fig:Omeganoiseless}
\end{figure*}

\begin{figure*}
    \centering
    \subfigure[Total frequency band of Channels A and E]{\includegraphics[width=0.47\textwidth]{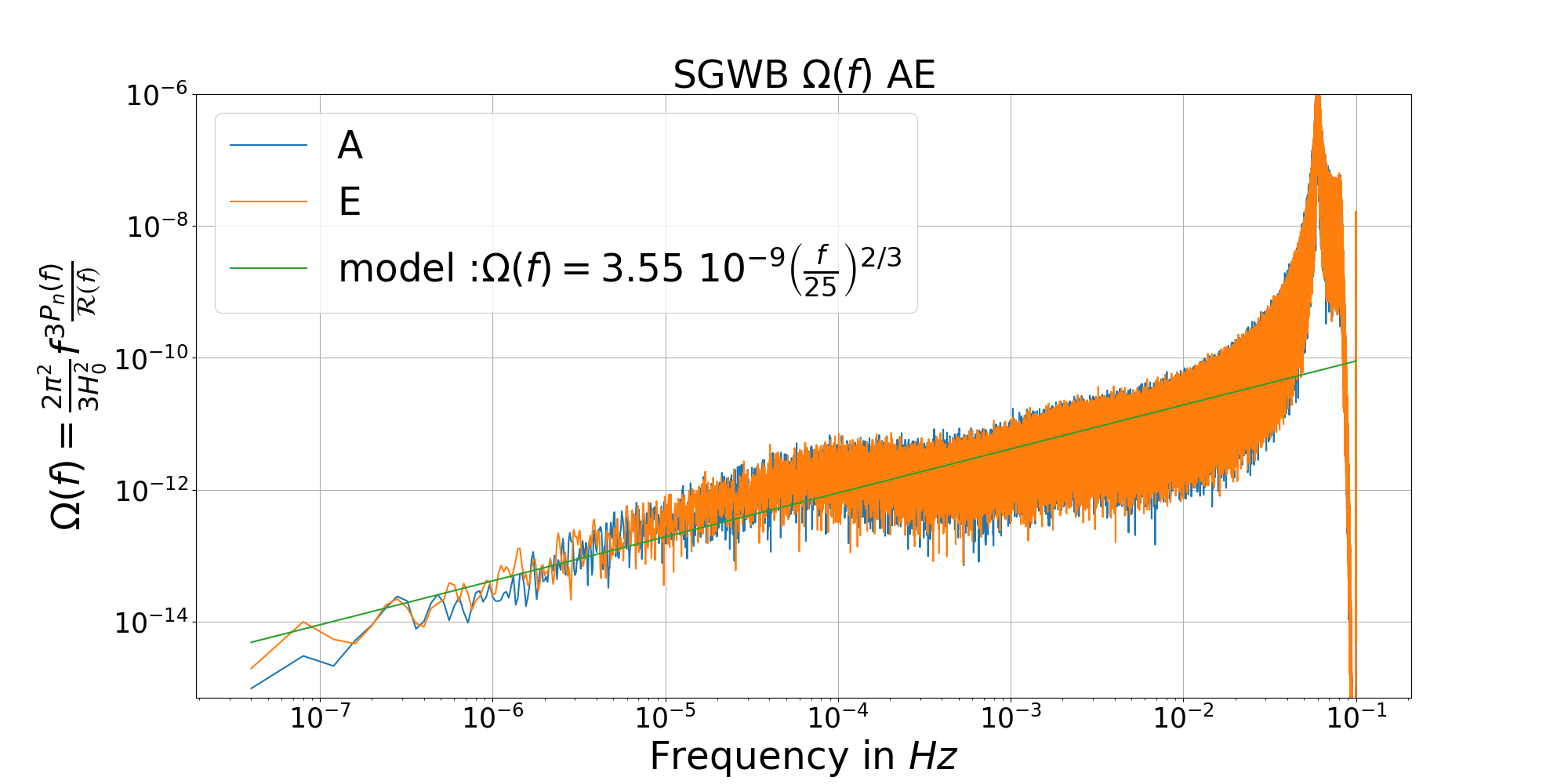}} 
    \subfigure[Reduce frequency band $2.15 \times 10^{-5}$ to $9.98 \times 10^{-3} \ \text{Hz}$ of Channels A and E]{\includegraphics[width=0.48\textwidth]{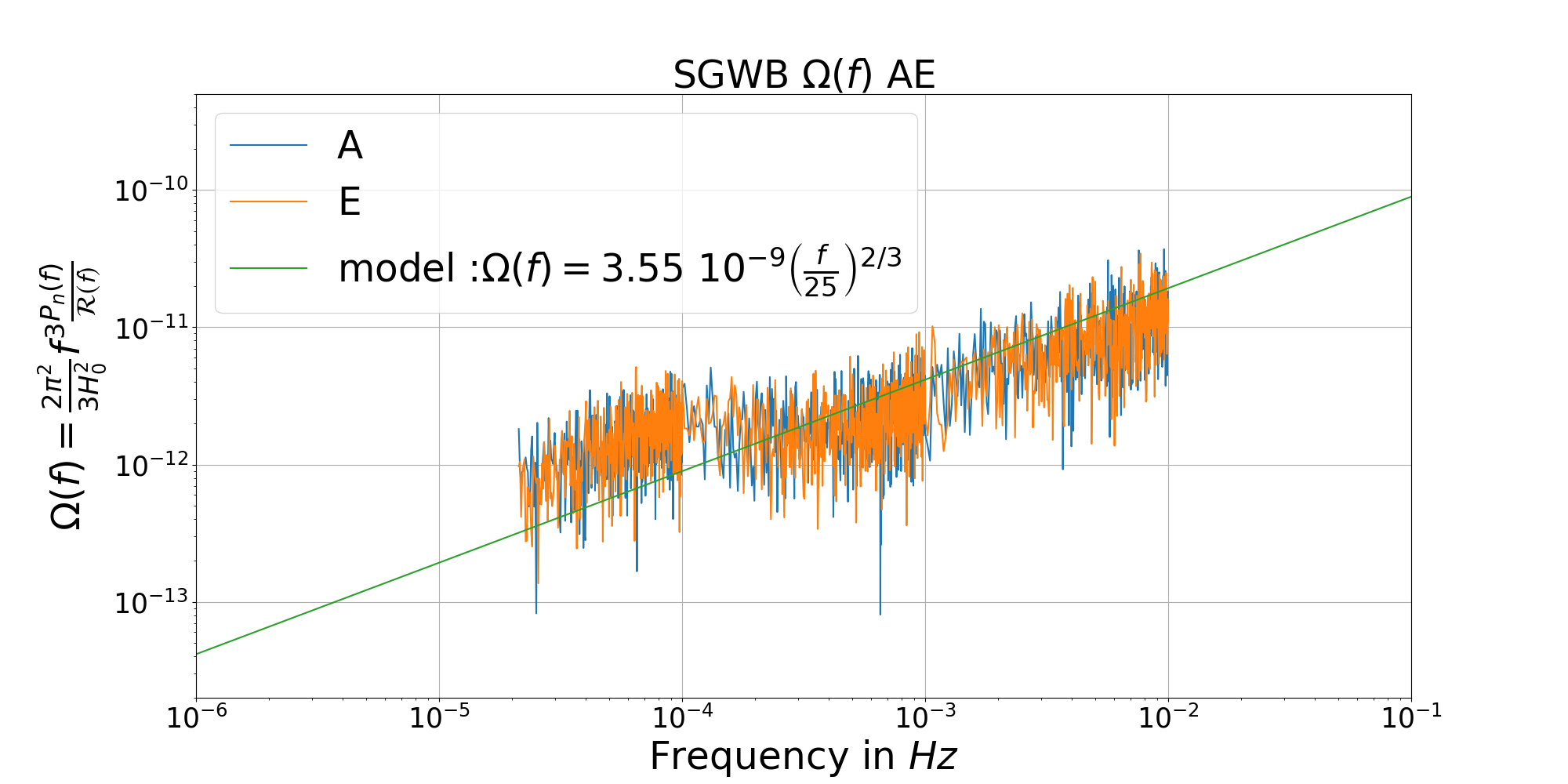}} 
    \caption{ Observations in channels [$A$,$E$] of the spectral energy density of the SGWB from astrophysical background $\Omega_{GW}(f)$ of the MLDC for the noisy channel, Eq.~\ref{eq:R_noisely}. (a) Total frequency band of Channels $A$ and $E$. (b) Reduced frequency band $2.15 \times 10^{-5}$ to $9.98 \times 10^{-3} \ \text{Hz}$ of Channels $A$ and $E$.}\label{fig:Omeganoisely}
\end{figure*}

We use two different response functions for the MLDC data, one system of equations for the noiseless data Eq.~\ref{eq:R_noiseless}, and one for the noisy data Eq.~\ref{eq:R_noisely} 
\begin{equation}\label{eq:R_noiseless}
    \left\{
\begin{array}{l}
    R_A(f) = R_{AA}(f)\frac{16}{9}\frac{2}{\pi}\left(\frac{f}{f_*}\right)^4 \sin^{-2}(f/f_*) \\
   R_E(f) = R_{EE}(f)\frac{16}{7}\frac{2}{\pi}\left(\frac{f}{f_*}\right)^4 \sin^{-2}(f/f_*)
\end{array}
\right.
\end{equation}
with $R_{II}$ given in~\cite{Adams:2010vc}, $f_* = \frac{c}{2 \pi L}$, and
\begin{equation}
    \begin{split}
        R_{AA}(f) &= R_{EE}(f) =  4 {\sin}^2\left(\frac{f}{f_*}\right)\Bigg[ \frac{3}{10} + \frac{169}{1680} \left(\frac{f}{f_*}\right)^2 \\ 
        &+\frac{85}{6048} \left(\frac{f}{f_*}\right)^4 - \frac{178273}{15667200} \left(\frac{f}{f_*}\right)^6 \\&
        + \frac{19121}{2476656000} \left(\frac{f}{f_*}\right)^8 \Bigg]
    \end{split}
\end{equation}
\begin{equation}
    R_I(f) = \frac{S_{II}(f)L}{3 c S_p} \left [\frac{36}{10}\frac{f}{f_*} \sin^{-2}(f/f_*) \right ]^2
\label{eq:R_noisely}
\end{equation}
where $S_{II}(f) = 8 \sin^{2}\left(\frac{f}{f_*}\right) \Bigg [4 S_a \bigg(1 + \cos\left(\frac{f}{f_*}\right) + \cos^2 \left(\frac{f}{f_*}\right) \bigg) + S_p \bigg(2+ \cos\left(\frac{f}{f_*}\right) \bigg)\Bigg]$ defined in~\cite{Romano2017} with $S_a= \frac{9\times 10^{-50}}{(2 \pi f)^4 }\bigg(1 + \left(\frac{10^{-4}}{f}\right)^2\bigg) $, $S_p = 4 . 10^{-42} \text{ Hz}^{-1}$ and $f_* = \frac{c}{2 \pi L}$. 
The energy spectral  density of the astrophysical background from the MLDC is a power law according to the documentation of the LISA Data Challenge Manual~\cite{LDCM} given by $\Omega_{GW}(f) = 3.55 \times 10^{-9}\left(\frac{f}{25 \ \text{Hz}}\right)^{2/3}$. 
Fig.~\ref{fig:Omeganoiseless} and Fig.~\ref{fig:Omeganoisely} show the energy periodogram $\widehat{\Omega}_{GW,I}(f)=\frac{2 \pi^2 }{3 H_0^2}f^3 \frac{I_n(f)}{R_I(f)}$ for channel $A$ in blue  and for channel $E$ in orange. The green curve is the power law model with the parameters $(\Omega_{\alpha}, f_{ref}, \alpha)$ with $\Omega_{GW} = \Omega_{\alpha}\left(\frac{f}{f_{ref}}\right)^{\alpha}$ from the MLDC documentation. The data at high frequency cannot be used because the transformation of the Eqs.~\ref{eq:R_noiseless} and \ref{eq:R_noisely} are valid for low-frequency. We use the frequency band  $[2.15 \times 10^{-5},  9.98 \times 10^{-3}] \ \text{Hz}$.

\subsection{Uncertainty of the Cosmological Component
$\Omega_{0}$ from the Adaptive Markov chains Monte Carlo (A-MCMC) } \label{sc:fisher+MCMC}
According to the Sec.~\ref{sec:Cramer}, one can calculate the uncertainty of the estimation of parameter $\Omega_0$ (the cosmological amplitude of the Spectral Energy Density), namely $\frac{\Delta \Omega_0}{\Omega_0}$.  To estimate this quantity from the Fisher Information, we use the formulae given in Sec.~\ref{spectral} and the inverse matrix of the Fisher Information (blue line in Fig.~ \ref{fig:Fisher2AET}). 

Not surprisingly we can predict a better separability (uncertainty is less) for high values of the cosmological background. 
The uncertainty can be calculated independently with the A-MCMC calculation: 
\begin{equation}
\frac{\Delta \Omega_0}{\Omega_0} = \frac{\sigma_{\Omega_0}}{\Omega_0}
\end{equation}
This ratio is calculated and represented as the scatter points on Fig.~\ref{fig:Fisher2AET}. We can also estimate the error of the uncertainty estimation (see Eq.~\ref{eq:errorbar}) from the estimation of the full width at half maximum of the posteriors distributions. The uncertainties (from the A-MCMC) are given by:
\begin{equation}
\label{eq:errorbar}
\left\{
\begin{array}{l}
    {Error}_{+,I} = \frac{\sigma_{\Omega_0}}{\left|\Omega_0-\sigma_{\Omega_0}\right|} \\
    {Error}_{-,I} = \frac{\sigma_{\Omega_0}}{\left|\Omega_0+\sigma_{\Omega_0}\right|}
\end{array}
\right.
\end{equation}

\begin{figure*}[t]
    \centering
    \includegraphics[height= 7cm]{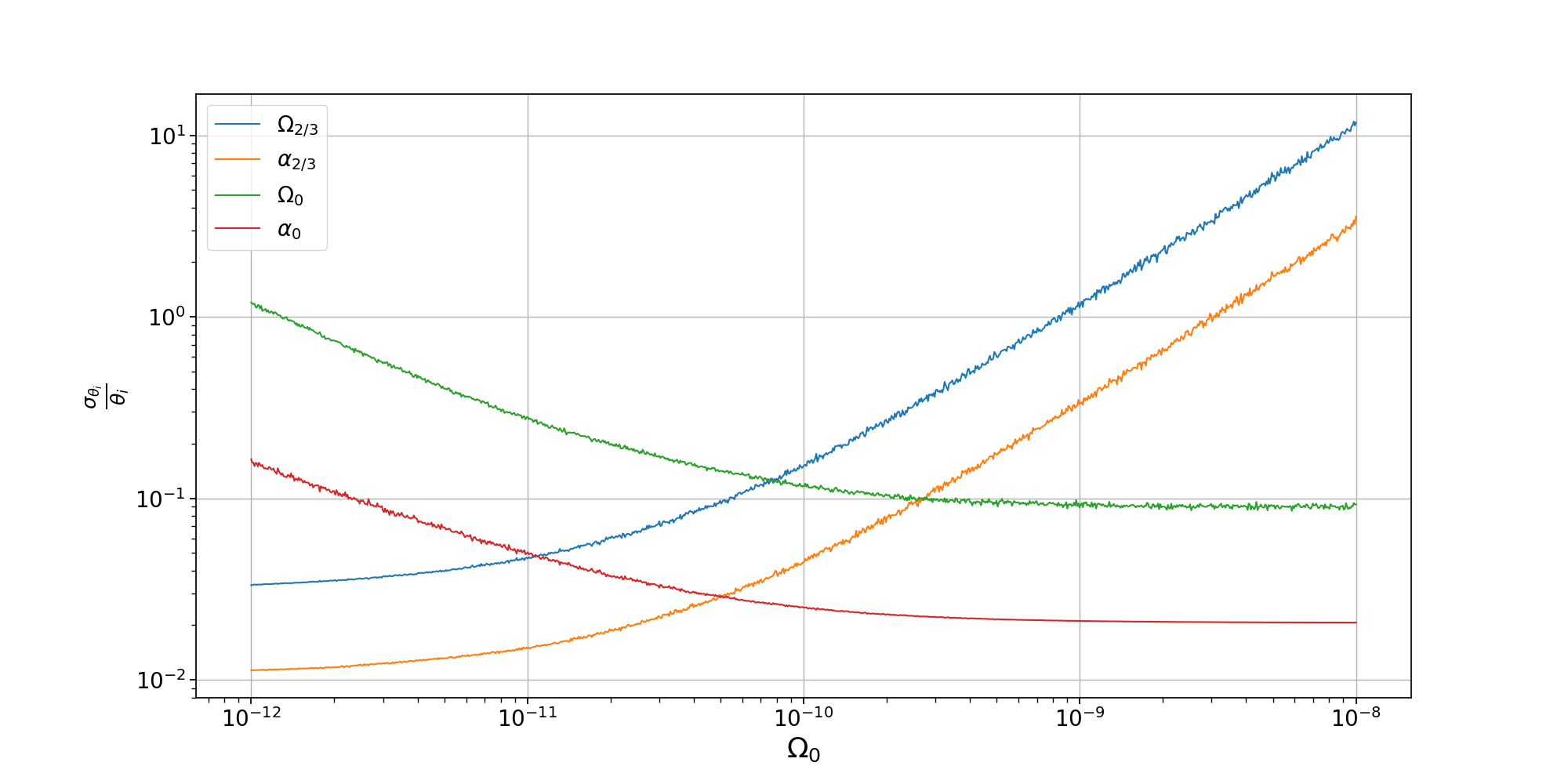}
    \caption{Evolution of the relative uncertainties for the estimation of the parameters $[\Omega_0, \alpha_0, \Omega_{2/3}, \alpha_{2/3}]$ versus the cosmological background amplitude $\Omega_0$. The precision for estimating the parameters is affected by the value of the cosmological amplitude $\Omega_0$. We use $\Omega_{2/3} = 3.55 \times 10^{-9}$, $\alpha_{2/3} = \frac{2}{3}$ and $\alpha_0 = 0$}
    \label{fig:FisheriiAE}
\end{figure*}

\section{Stochastic gravitational-wave background fitting with Adaptive Markov chain Monte-Carlo using the channel $T$ and the two science channels $A$ and $E$}\label{sc:cov}
\label{sec:AET}

In this section we consider the null channel $T$ and the science channels $A$ and $E$. We assume that the observation of the noise in channel $T$ informs us of the noise in channels $A$ and $E$. We follow the formalism of Smith and Caldwell~\cite{PhysRevD.100.104055}. 

We can simulate the noise and SGWB in frequency domain.
\begin{equation}\label{eq:modelPSDs}
\left\{
\begin{array}{l}
    PSD_A = S_A + N_A \\
    PSD_E = S_E + N_E \\
  PSD_T = N_T
\end{array}
\right.
\end{equation}
With $S_A(f) = S_E(f) =\frac{3H_0^2}{4 \pi^2} \frac{\Omega_{GW,\alpha}\left(\frac{f}{f_{ref}}\right)^{\alpha}\mathcal{R}(f)}{f^3}$, $f_{ref}=25 \ \text{Hz}$, the noise components $N_A(f)=N_E(f)$ and $N_T(f)$ can be written as:
\begin{equation}
\left\{
\begin{array}{l}
    N_A = N_1 - N_2 \\
    N_T = N_1 + 2 N_2 
\end{array}
\right.
\end{equation}
with 
\begin{equation}
\left\{
\begin{array}{l}
    N_1(f) = \left(4 S_s(f) + 8\left( 1 + \cos^2\left(\frac{f}{f_*}\right)\right) S_a(f)\right)|W(f)|^2 \\
    N_2(f) = -\left(2 S_s(f) + 8 S_a(f)\right)\cos\left(\frac{f}{f_*}\right)|W(f)|^2
\end{array}
\right.
\end{equation}
$W(f) = 1 - e^{-\frac{2if}{f_*}}$ and 
\begin{equation}
\left\{
\begin{array}{l}
    S_s(f) = N_{Pos} \\
    S_a(f) = \frac{N_{acc}}{(2 \pi f)^4}\left( 1 + \left(\frac{0.4 \text{ mHz}}{f} \right)^2 \right)
\end{array}
\right.
\end{equation}
The LISA noise budget is given from the \href{https://atrium.in2p3.fr/nuxeo/nxdoc/default/f5a78d3e-9e19-47a5-aa11-51c81d370f5f/view_documents}{LISA Science Requirement} Document~\cite{LSR}. To create the data for our example, we use an acceleration noise of $N_{acc} = 1.44 \times 10^{-48} \ \text{s}^{-4}\text{Hz}^{-1}$ and the optical path-length fluctuation $N_{Pos} = 3.6 \times 10^{-41} \ \text{Hz}^{-1}$. 
We can estimate the magnitude of the noise from the channel $T$. 
One should note the importance of using the channel $T$ to estimate the noise in the channels $A$ and $E$, as it is then possible to parameterize an A-MCMC of six parameters, $\theta = (N_{acc},N_{Pos},\Omega_{2/3}, \alpha_{2/3},\Omega_0, \alpha_0)$.  We can also calculate the propagation of uncertainties for the power spectral densities with the partial derivative method. As such, we can estimate the error on the measurement realized by a fit of the parameters $\theta$, $dPSD_I = \sqrt{\sum_{\theta} \left(\frac{\partial PSD_I} {\partial \theta}\right)^2 d\theta^2}$. We then obtain for two SGWBs
$\Omega_{astro}(f) = \Omega_{2/3}\left(\frac{f}{f_{ref}}\right)^{2/3}$, $\Omega_{cosmo}(f) = \Omega_0\left(\frac{f}{f_{ref}}\right)^0$,  
\begin{equation}\label{eq:errorPSD}
\left\{
\begin{array}{l}
    \begin{split}
        dPSD_I &= \Big[N_I(0, dN_{acc},f)^2 + N_I(dN_{pos}, 0,f)^2 \\
        &+ S_I(\Omega_{2/3}, \alpha_{2/3},\Omega_0, \alpha_0,f)^2 \Big(d\Omega_0^2 + d\Omega_{2/3}^2\\
        &+ \ln{\left(\frac{f}{f_{ref}}\right)^2\Big(\Omega_{2/3}^2 d\alpha_{2/3}^2 + \Omega_0^2 d\alpha_0^2\Big) \Big)}\Big]^{1/2}
    \end{split}\\
    dPSD_T = \Big[N_T(0, dN_{acc},f)^2 + N_T(dN_{pos}, 0,f)^2\Big]^{1/2}
\end{array}
\right.
\end{equation}
with $\{dN_{acc}, dN_{pos}, d\Omega_{astro}, d\alpha_{astro}, d\Omega_{cosmo}, d\alpha_{cosmo} \}$ being the positive error estimations of the parameters; $I = A,E$. We take 1 $\sigma$ for the posterior distributions.  
We can also estimate the error of the power spectral density fit using the MCMC chains to produce the error. With the MCMC chains we can calculate a histogram of $PSD_I(f)$ at each frequency. For each histogram we compute the $68\%$ credible band. This method is similar to that of BayesWave; see Figure 7 of \cite{2020CQGra..37e5002A}. The two methods produce the same error bands, but we need to assume that the posterior distributions are Gaussian. The quadratic sum of partial errors calculation yields a good estimation of error from MCMC chains if the posterior distributions of the chains are Gaussian.    

We can calculate the covariance matrix:
\begin{equation}
    <PSD_I(f),PSD_J(f)> = \mathcal{C}_{I,J}(\theta,f) 
\end{equation}
with $I,J = [A,E,T]$.
As such, it is possible to parameterize an A-MCMC with six parameters: $\theta = (N_{acc},N_{Pos},\Omega_{GW\alpha},\alpha)$.
We can calculate the covariance matrix of $(\tilde{d}_A(f),\tilde{d}_E(f),\tilde{d}_T(f))$
\begin{equation}
     \mathcal{C}(\theta,f) =
     \left(
     \begin{array}{ccc}
      S_A + N_A & 0 & 0  \\
      0 & S_E + N_E & 0 \\
      0 & 0 & N_T \\
     \end{array}
     \right)
\end{equation}
\begin{equation}
     \mathcal{C}^{-1}(\theta, f) = K
     \left(
     \begin{array}{ccc}
      (S_A + N_A)^{-1} & 0 & 0  \\
      0 &  (S_E + N_E)^{-1} & 0  \\
      0 & 0 &  N_T^{-1} \\
     \end{array}
     \right)
 \end{equation}
 and $K(f_k) = det(\mathcal{C}) = \frac{1}{(S_A + N_A)(S_E + N_E)N_T}$.
 We use the definition of the Whittle likelihood from~\cite{Romano2017}, and the log-likelihood is:
\begin{equation} 
\begin{split}
     \mathcal{L}(\textbf{d}|\theta) & =  -\frac{1}{2} \sum_{k=0}^N \Bigg[ \sum_{I,J = [A,E,T]} \left( \sqrt{d_I(f)} \left(\mathcal{C}^{-1}\right)_{IJ} \sqrt{d_J(f)} \right) \\
     &+ \ln\left(2\pi K(f_k) \right) \Bigg] \\
    &=  -\frac{1}{2} \sum_{k=0}^N \Bigg[ \frac{d_A^2}{S_A+N_A}  + \frac{d_E^2}{S_E+N_E} + \frac{d_T^2}{N_T} \\
    &+  \ln\left(8\pi^3 (S_A+NA)(S_E+N_E)N_T \right) \Bigg] 
\end{split}
\end{equation}
\begin{equation} 
    \begin{split}
        F_{ab} &=   \frac{1}{2} \mathrm{Tr}\left(\mathcal{C}^{-1}\frac{\partial \mathcal{C}} {\partial \theta_a} \mathcal{C}^{-1} \frac{\partial \mathcal{C}}{\partial\theta_b}  \right) \\
        &=  M \sum_{k=0}^{N} \Bigg[\frac{\frac{\partial (S_A+N_A)} {\partial \theta_a}\frac{\partial (S_A+N_A)} {\partial \theta_b}}{2(S_A+N_A)^2} \\
        &+ \frac{\frac{\partial (S_E+N_E)} {\partial \theta_a}\frac{\partial (S_E+N_E)} {\partial \theta_b}}{2(S_E+N_E)^2} + \frac{\frac{\partial N_T} {\partial \theta_a}\frac{\partial N_T} {\partial \theta_b}}{2N_T^2}\Bigg]
    \end{split}
\end{equation}
with $M=Df_b$ ($D$ is the time duration of the LISA mission and $f_b$ the highest frequency of interest in the LISA band).
If we have the channel $T$ as zero and we consider the two science channels $A$ and $E$ as independent, we obtain:
\begin{equation}
    F_{ab} = M \frac{1}{2} \sum_{I=A,E} \sum_{k=0}^N \frac{\frac{\partial S_I(f)+ N_I(f)} {\partial \theta_a}\frac{\partial S_I(f)+ N_I(f)} {\partial \theta_b}}{\left(S_I(f)+ N_I(f)\right)^2} 
\end{equation}
We have a comparable result given in \cite{PhysRevD.100.104055}, the inverse of the Fisher Information matrix on the diagonal gives the uncertainties of the estimation of the parameters. We see the importance to estimate the "noise" channel $T$ for the estimation of the SGWB.  

In Fig.~\ref{fig:FisheriiAE} we display the influence of the precision on the fitted parameter versus the value of the cosmological background $\Omega_0$. Obviously, we understand that if the astrophysical background is large it will be harder to measure the cosmological background with high precision.

We have also conducted an A-MCMC study with 6 parameters: 2 for the noise channel $T$, 2 for the astrophysical background, and 2 for the cosmological background. We use the data from the two science channels, $A$ and $E$, along with channel $T$. Given the magnitude level of the LISA noise budget from the LISA Science Requirements Document \cite{LSR}, we use the acceleration noise $N_{acc} = 1.44 \times 10^{-48} \ \text{s}^{-4}\text{Hz}^{-1}$ and the optical path-length fluctuation $N_{Pos} = 3.6 \times 10^{-41} \ \text{Hz}^{-1}$.
 We make the assumption that the data in Channel A and T are independent. The noise in both channels depend  on the two parameters $N_{pos}$ and $N_{acc}$. 
 We aim to estimate the SGWB and noise parameters simultaneously using data from both channels $A,E$ and $T$ via our A-MCMC algorithm. Using the additional data from channel $T$ will yield a more efficient estimation procedure and a gain in precision of parameter estimates than using the data from channels $A, E$ only. 
For four different magnitudes of the astrophysical SGWB, we conduct A-MCMC runs with different values for the amplitude of the cosmological background; see Table~\ref{table:resultA+E+T}). The A-MCMC is characterized by $\beta = 0.01$, $N = 4 \ 000 \ 000$ (see Section~\ref{sec:adaptive}) and we use 2 000 samples to estimate the co-variance matrix. We use log uniform priors with 10 magnitude intervals for the 2 noise channel parameters $[N_{Opt},N_{Acc}]$ and for the two background amplitudes $[\Omega_{cosmo}, \Omega_{astro}]$, a uniform prior for the slope between $-0.4$ and $0.4$ for the cosmological slope $\alpha_{cosmo}$, and a uniform prior between $0.27$ and $1.07$ for the astrophysical slope $\alpha_{astro}$.

We note for comparison purposes the results given in~\cite{PhysRevD.100.104055} where the diagonal elements of the inverse of the Fisher Information $F_{ab}$ provide  the uncertainties of the respective parameter estimates. 
The Fisher Information Matrix is a Block matrix. Indeed, we have a $6 \times 6$ matrix, assuming the parameters to be independent. We can thus distinguish two independent types, the first coming from derivatives related to the noise of LISA this generates a $ 2 \times2 $ matrix, $N_{2 \times2}$. The second type corresponds to a $ 4 \times4 $ matrix giving the derivatives linked to the SGWB, $S_{4 \times4}$. This second matrix is the same as the one calculated in the Sec.~\ref{sc:FisherInfo}. So we have:
\begin{equation}
        F_{ab} = \left[ 
    \begin{array}{c|c} 
      N_{2 \times2} & 0 \\ 
      \hline 
      0 & S_{4 \times4}
    \end{array} 
    \right] 
\end{equation}
\begin{figure*}[t]
    \centering
    \includegraphics[height= 18cm]{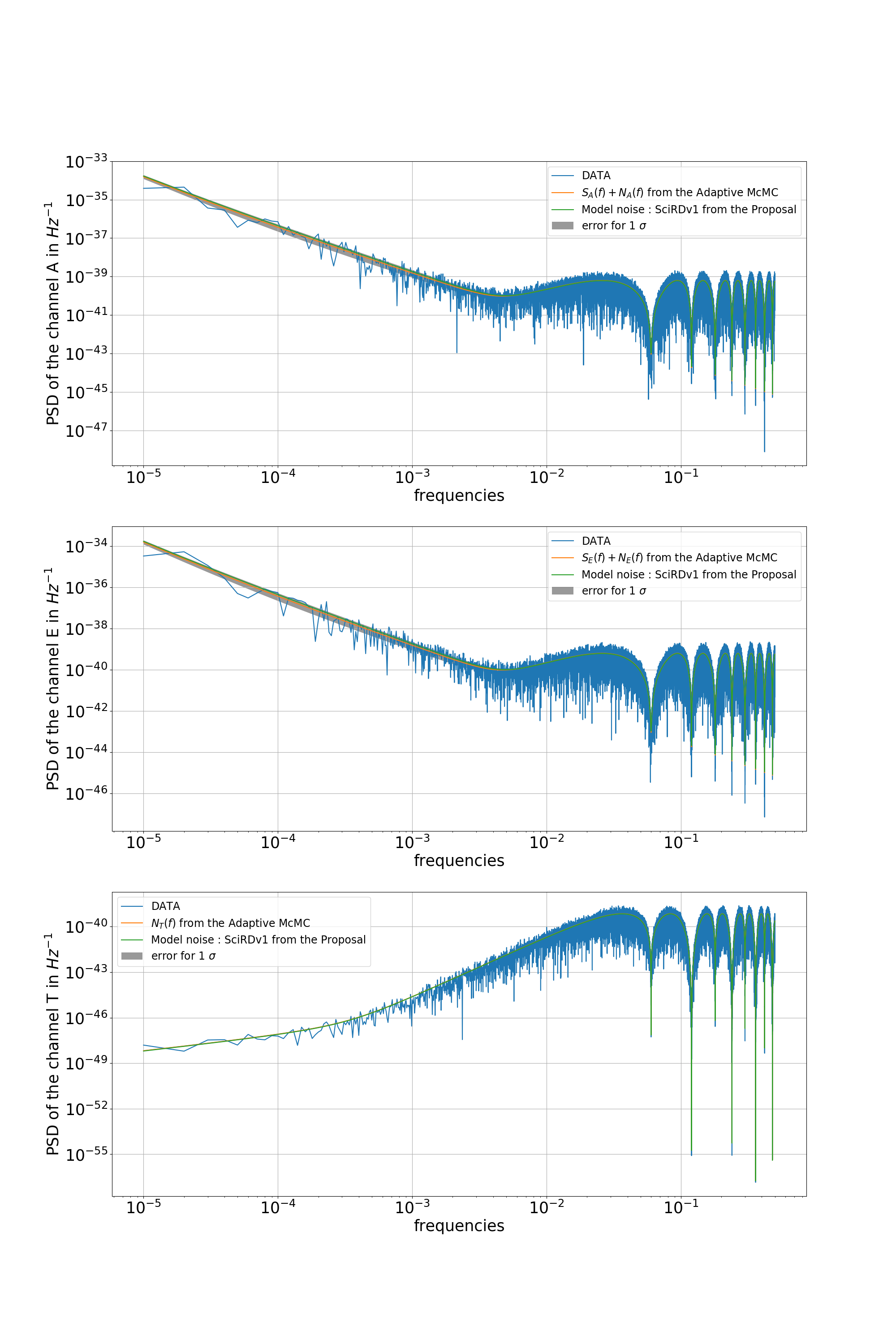}
    \caption{Power spectral density of the channels $A$, $E$ and $T$  from the LISA noise model~\cite{PhysRevD.100.104055} and an astrophysical SGWB ($\Omega_{2/3} = 3.55 \times 10^{-9}$ at 25 Hz). The figures show the power spectral densities: channel $A$ top, $E$ middle, and $T$ bottom. The parameters are from the proposal~\cite{LSR}. The orange line is the LISA noise model from~\cite{PhysRevD.100.104055}, in green the values from the A-MCMC, and in grey the 1 $\sigma$ error.}
    \label{fig_covMCMC}
\end{figure*}

\begin{figure*}[t]
    \centering
    \includegraphics[height= 12cm]{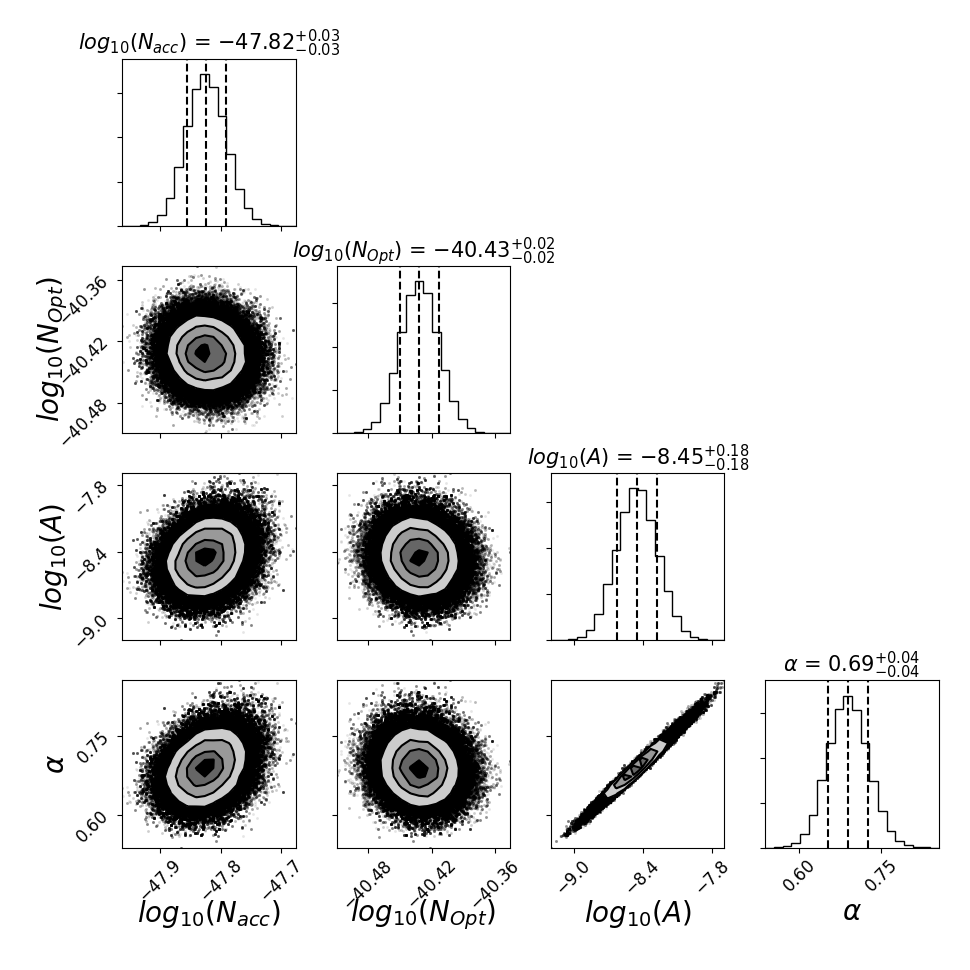}
    \caption{Corner plot for the A-MCMC using the channels $A$, $E$ and $T$. The results are for the two magnitudes for the LISA noise model from the proposal~\cite{LSR}, and a single SGWB (amplitude and spectral slope). The vertical dashed lines on the posterior distribution represent from left to right the quantiles $[16\%,\ 50\%,\ 84\% ]$. The {\it true} values for the parameters are $\theta = (N_{acc},N_{Pos},\Omega_{GW\alpha},\alpha) = \bigg(1.44 \times 10^{-48} \ \text{s}^{-4} \text{Hz}^{-1}$, $3.6 \times 10^{-41} \ \text{Hz}^{-1}$, $3.55 \times 10^{-9}$,~$\frac{2}{3}\bigg)$.}
    \label{fig_corncovMCMC}
\end{figure*}

In Fig.~\ref{fig_covMCMC}, the blue line is the data for $\theta = (N_{acc},N_{Pos},\Omega_{GW\alpha},\alpha) = \bigg(1.44 \times 10^{-48} \ \text{s}^{-4} \text{Hz}^{-1}$, $3.6 \times 10^{-41} \ \text{Hz}^{-1}$, $3.55 \times 10^{-9}$,~$\frac{2}{3}\bigg)$. The data are simulated with the LISA noise model of the Eq.~\ref{eq:modelPSDs} with a SGWB from binaries origin. 
The green line is the LISA noise model from \cite{PhysRevD.100.104055}. 
The A-MCMC is characterized by $\beta = 0.01$, $N = 1 \ 000 \ 000$ (see Sec.~\ref{sec:adaptive}) and we use 2 000 samples to estimate the co-variance matrix. We use log uniform priors with 10 magnitude intervals for the three first parameters and a uniform prior for the slope between $-\frac{4}{3}$ and $\frac{8}{3}$. The orange line in Fig.~\ref{fig_covMCMC} displays the result of the A-MCMC, and in grey the error for 1 $\sigma$. Fig.~\ref{fig_corncovMCMC} displays the corner plot from the A-MCMC; the posterior distributions are well approximated by Gaussian distributions. We have evidence of good fits. The estimation of the noise level magnitudes from the parametric estimation yields a positive result because we have the possibility to fit the background with the noise level throughout the frequency domain; it is also possible to have a very efficient estimation of the different noise components thanks to the signal $T$ being devoid of a science signal source.

The advantage of 2 science channels, $A$ and $E$, as opposed to one, $A$ or $E$, is a factor of $\sqrt{2}$ for the error estimation, and hence the overall sensitivity. Indeed, the error of the cosmological amplitude is given by the coefficient $(\Omega_0,\Omega_0)$ of the square root of the inverse of the Fisher Information matrix. We have for one channel ($A$ or $E$), ${\Delta \Omega_0}_{(A \ \text{or } E)} = \sqrt{{F_{\Omega_0,\Omega_0}^{-1}}_{(A \ \text{or } E)}}$. For a combination of $A$ and $E$, we have $\Delta {\Omega_0}_{(A \ \text{and } E)}  = \frac{{\Delta \Omega_0}_{A \ \text{or } E)}}{\sqrt{2}}$, because the two channels respond identically. If we modeled the spectrum of the two Channel $A$ and $E$ as the same think we would have ${F_{a,b}}_{(A \ \text{and } E)} = 2 {F_{a,b}}_{(A \ \text{or } E)}$.

\begin{table*}[!b]
\begin{ruledtabular}
\begin{tabular}{c|cccc|cccc}
  Input &                          \multicolumn{4}{c|}{Values of the A-MCMC}                   &                      \multicolumn{4}{c}{errors ($\sigma$)}                        \\ \hline
                \backslashbox{$\Omega_0$}{$\Omega_{Astro}$}  
                & $3.55 \times 10^{-8}$   & $3.55 \times 10^{-9}$   & $1.8 \times 10^{-9}$    & $3.55 \times 10^{-10}$  & $3.55 \times 10^{-8}$   & $3.55 \times 10^{-9}$   & $1.8 \times 10^{-9}$    & $3.55 \times 10^{-10}$  \\ \hline
  $1. \times 10^{-8}$  & $1.011 \times 10^{-8}$  & $9.982 \times 10^{-9}$  & $9.987 \times 10^{-9}$  & $9.992 \times 10^{-9}$  & $3.395 \times 10^{-10}$ & $3.057 \times 10^{-10}$ & $3.106 \times 10^{-10}$ & $2.588 \times 10^{-10}$ \\ 
  $5. \times 10^{-9}$  & $5.014 \times 10^{-9}$  & $4.971 \times 10^{-9}$  & $5.007 \times 10^{-9}$  & $4.960 \times 10^{-9}$  & $1.754 \times 10^{-10}$ & $1.464 \times 10^{-10}$ & $1.506 \times 10^{-10}$ & $1.462 \times 10^{-10}$ \\ 
  $2. \times 10^{-9}$  & $2.005 \times 10^{-9}$  & $1.984 \times 10^{-9}$  & $2.007 \times 10^{-9}$  & $2.083 \times 10^{-9}$  & $7.481 \times 10^{-11}$ & $5.600 \times 10^{-11}$ & $6.588 \times 10^{-11}$ & $5.492 \times 10^{-11}$ \\ 
  $1. \times 10^{-9}$  & $9.972 \times 10^{-10}$ & $1.008 \times 10^{-9}$  & $1.046 \times 10^{-9}$  & $1.046 \times 10^{-9}$  & $4.480 \times 10^{-11}$ & $2.828 \times 10^{-11}$ & $3.196 \times 10^{-11}$ & $3.196 \times 10^{-11}$ \\ 
  $5. \times 10^{-10}$ & $4.965 \times 10^{-10}$ & $4.975 \times 10^{-10}$ & $5.076 \times 10^{-10}$ & $4.956 \times 10^{-10}$ & $2.529 \times 10^{-11}$ & $1.497 \times 10^{-11}$ & $1.703 \times 10^{-11}$ & $1.385 \times 10^{-11}$ \\ 
  $2. \times 10^{-10}$ & $2.002 \times 10^{-10}$ & $1.984 \times 10^{-10}$ & $1.976 \times 10^{-10}$ & $1.976 \times 10^{-10}$ & $1.394 \times 10^{-11}$ & $6.647 \times 10^{-11}$ & $8.251 \times 10^{-12}$ & $5.157 \times 10^{-11}$ \\ 
  $1. \times 10^{-10}$ & $9.981 \times 10^{-11}$ & $1.065 \times 10^{-10}$ & $9.941 \times 10^{-11}$ & $1.003 \times 10^{-10}$ & $9.228 \times 10^{-12}$ & $5.322 \times 10^{-12}$ & $4.050 \times 10^{-12}$ & $3.048 \times 10^{-12}$ \\ 
  $5. \times 10^{-11}$ & $5.013 \times 10^{-11}$ & $5.057 \times 10^{-11}$ & $5.058 \times 10^{-11}$ & $5.163 \times 10^{-11}$ & $7.078 \times 10^{-11}$ & $5.171 \times 10^{-12}$ & $2.879 \times 10^{-12}$ & $1.706 \times 10^{-12}$ \\ 
  $2. \times 10^{-11}$ & $2.006 \times 10^{-11}$ & $2.014 \times 10^{-11}$ & $1.989 \times 10^{-11}$ & $2.016 \times 10^{-11}$ & $5.389 \times 10^{-12}$ & $2.558 \times 10^{-12}$ & $1.130 \times 10^{-12}$ & $8.457 \times 10^{-13}$ \\ 
  $1. \times 10^{-11}$ & $1.001 \times 10^{-11}$ & $1.008 \times 10^{-11}$ & $1.002 \times 10^{-11}$ & $1.026 \times 10^{-11}$ & $4.269 \times 10^{-12}$ & $1.406 \times 10^{-12}$ & $5.902 \times 10^{-13}$ & $4.472 \times 10^{-13}$ \\ 
  $5. \times 10^{-12}$ & $5.011 \times 10^{-12}$ & $4.959 \times 10^{-12}$ & $5.001 \times 10^{-12}$ & $5.024 \times 10^{-12}$ & $3.583 \times 10^{-12}$ & $9.843 \times 10^{-13}$ & $4.526 \times 10^{-13}$ & $2.556 \times 10^{-13}$ \\ 
  $2. \times 10^{-12}$ & $2.196 \times 10^{-12}$ & $1.952 \times 10^{-12}$ & $1.948 \times 10^{-12}$ & $1.985 \times 10^{-12}$ & $3.001 \times 10^{-12}$ & $7.460 \times 10^{-13}$ & $3.190 \times 10^{-13}$ & $1.433 \times 10^{-13}$ \\ 
  $1. \times 10^{-12}$ & $1.019 \times 10^{-12}$ & $1.064 \times 10^{-12}$ & $9.936 \times 10^{-13}$ & $1.013 \times 10^{-12}$ & $2.155 \times 10^{-12}$ & $5.119 \times 10^{-13}$ & $2.233 \times 10^{-13}$ & $1.040 \times 10^{-13}$ \\  
   $1. \times 10^{-13}$ &  & $9.891 \times 10^{-14}$ & $1.040 \times 10^{-13}$ & $9.936 \times 10^{-14}$ &  & $2.002 \times 10^{-13}$ & $1.036 \times 10^{-13}$ & $4.054 \times 10^{-14}$ \\ 
\end{tabular}
\caption{Results of the A-MCMC runs with 6 parameters (2 for the LISA noise, 2 for the astrophysical background and 2 for the cosmological background). We use the data from the $A$, $E$ and $T$ channels. The four columns of values correspond to the output of 13  A-MCMC runs. The study is conducted using 4 values for the amplitude of the astrophysical background after 4 years of observation: $3.55 \times 10^{-8}$, $3.55 \times 10^{-9}$, $1.8 \times 10^{-9}$ and  $3.55 \times 10^{-10}$. And respectively, the same for the error columns. The error estimations come from the posteriors distributions.}
\label{table:resultA+E+T}
\end{ruledtabular}
\end{table*}

\begin{figure*}[t]
    \includegraphics[height= 8cm]{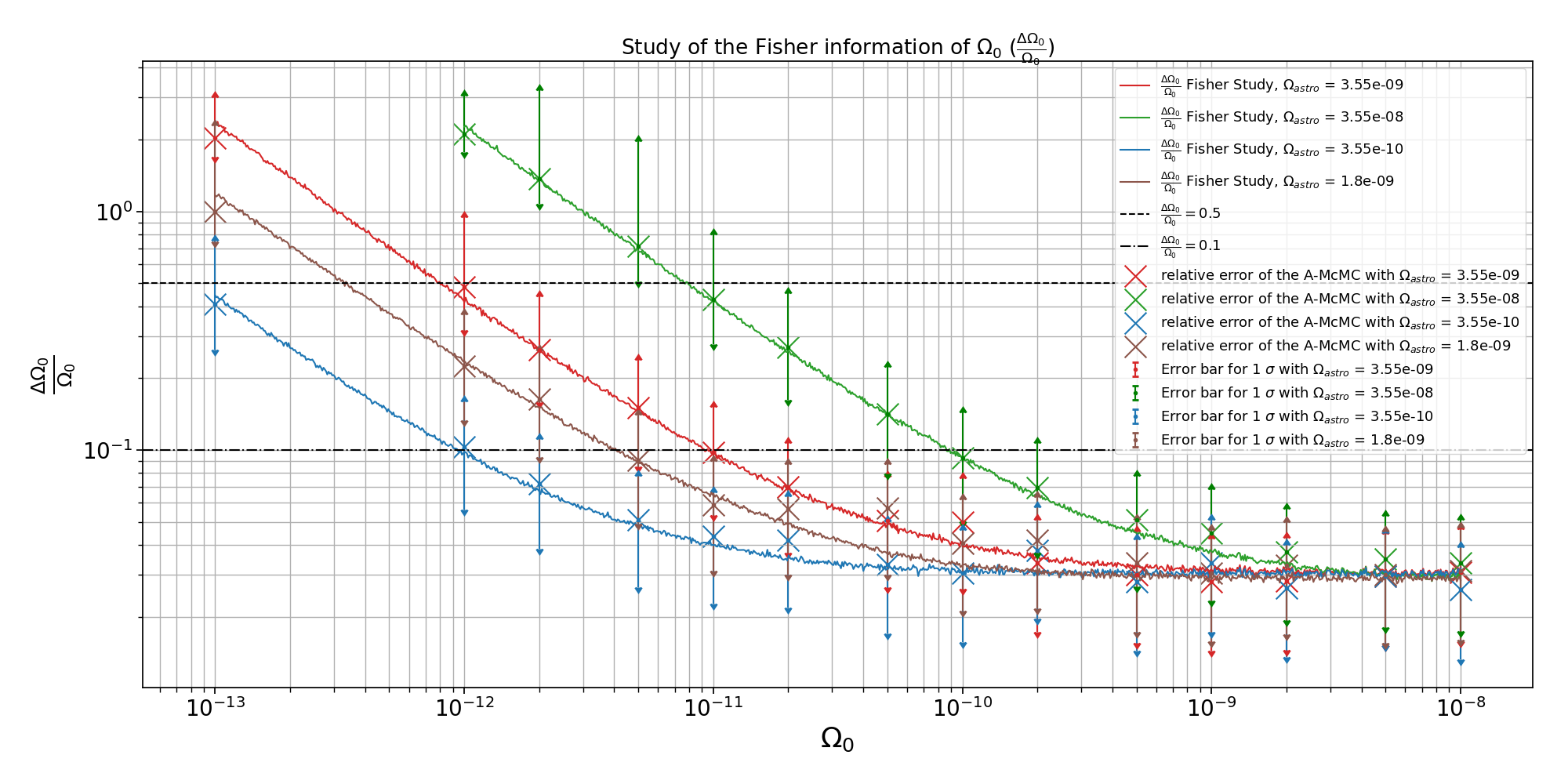}
    \caption{Uncertainty of the estimation of the parameter $\Omega_0$ (the spectral energy density of the cosmological SGWB) from the Fisher Information study (displayed with lines), and the parametric estimation from the A-MCMC (with the scatter points) for the channel $A$ and $E$, with the noise channel $T$. We conduct the study with different value for the astrophysical magnitude $\Omega_{astro}$. There are error bars for the four sets of A-MCMC runs; see Eq.~\ref{eq:errorbar}. The horizontal dashed line represents the error level of $50\%$. 
    TThis is the limit where it is possible to observe the cosmological SGWB. The semi-dashed line represents the $10\%$ error.}
    \label{fig:Fisher2AET}
\end{figure*}

\begin{figure*}[t]
    \centering
    \includegraphics[height= 14cm]{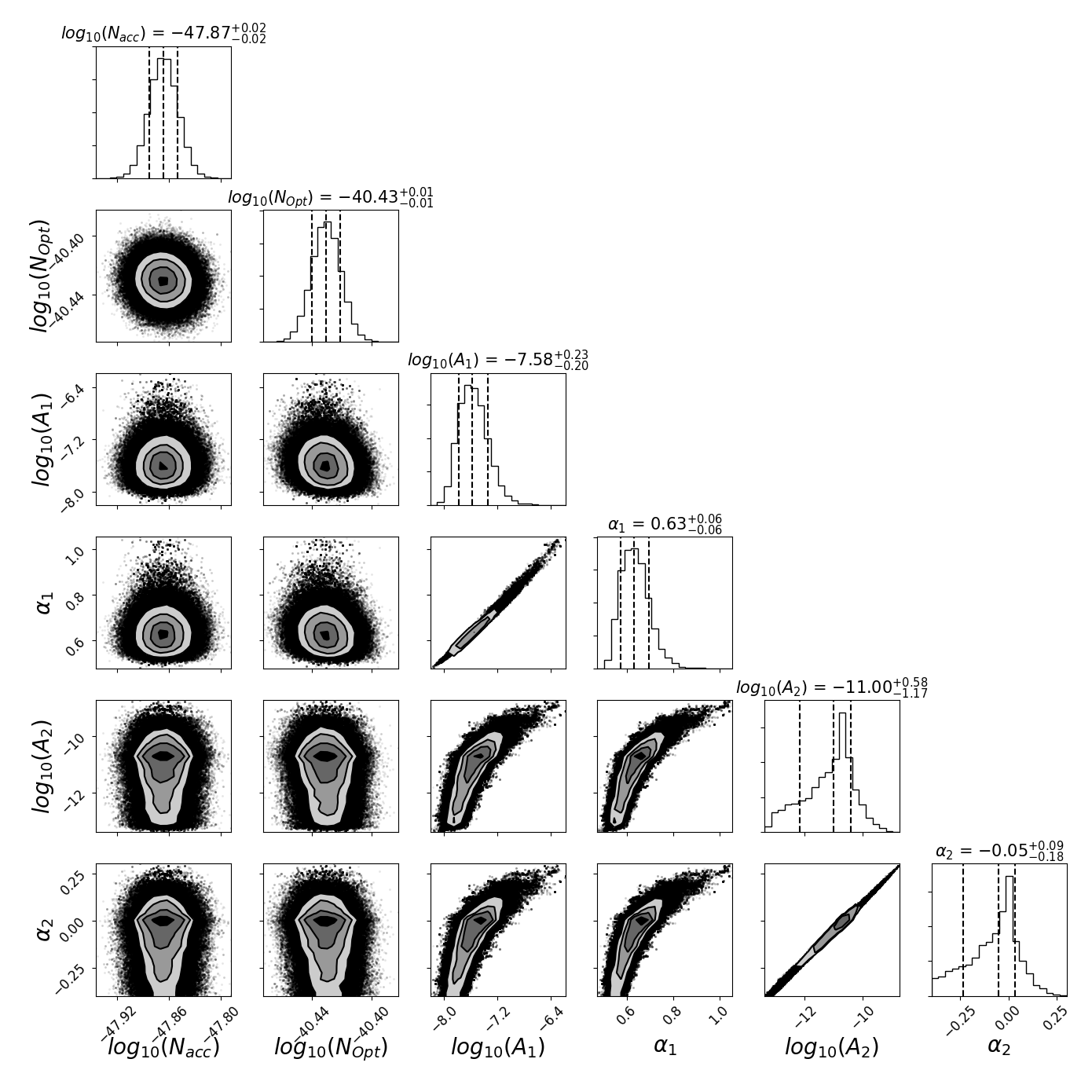}
    \caption{Corner plot giving the A-MCMC generated posterior distributions for a run with 6 parameters with $\Omega_{GW,Astro} = 3.55 \times 10^{-8}$ and  $\Omega_{GW,Cosmo} = 1 \times 10^{-11}$. The vertical dashed lines on the posterior distributions represent from left to right the quantiles $[16\%,\ 50\%,\ 84\% ]$.  This is from a run of using the data from channels $A$, $E$ and $T$. These results are presented in Table~\ref{table:resultA+E+T} and also in Fig.~\ref{fig:Fisher2AET}.}
    \label{fig:Corner6paramAET}
\end{figure*}

\begin{figure*}[t]
    \includegraphics[height= 14cm]{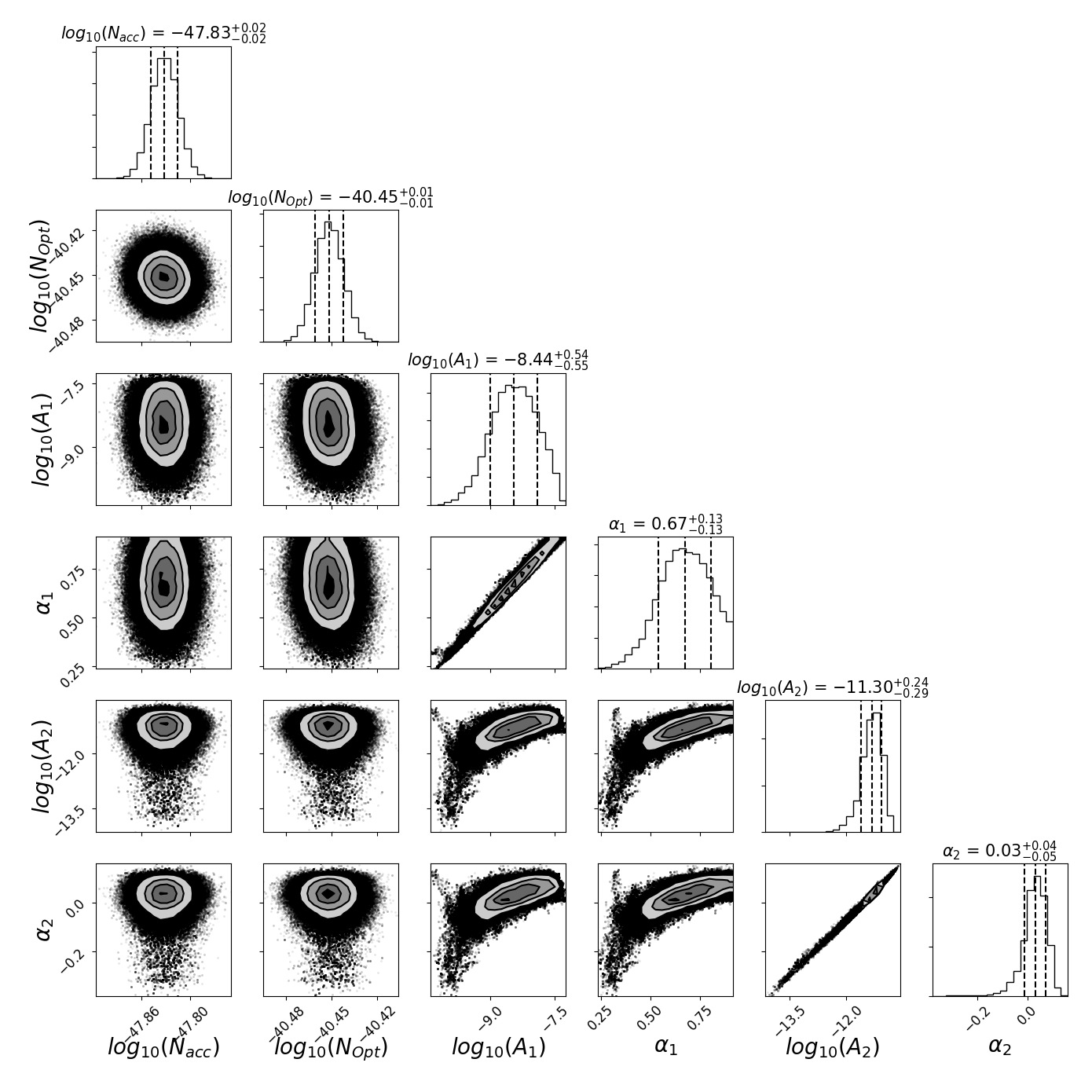}
    \caption{Corner plot giving the A-MCMC generated posterior distributions for a run of 6 parameters, with $\Omega_{GW,Astro} = 3.55 \times 10^{-9}$ and  $\Omega_{GW,Cosmo} = 5 \times 10^{-12}$. The vertical dashed lines on the posterior distribution represent from left to right the quantiles $[16\%,\ 50\%,\ 84\% ]$.  This is from a run using the data from channels $A$, $E$ and $T$. These results are presented in Table~\ref{table:resultA+E+T}, and also in Fig.~\ref{fig:Fisher2AET}.}
    \label{fig:Corner6param2AET}
\end{figure*}

Note that in the LISA observing band we have a ratio of $ \frac{\Omega_{astro}}{\Omega_{Cosmo}} = 5.29$ at 1 mHz and $1.15$ at $0.1$ mHz. The importance in being able to distinguish between two backgrounds is not the absolute amplitude of the background, but the ratio between the two backgrounds' magnitudes $\frac{\Omega_{astro}}{\Omega_{Cosmo}}$.  For a smaller ratio we can fit the cosmological background with less uncertainty.
From Fig.~\ref{fig:Fisher2AET}, we can separate the cosmological background from the astrophysical background with a magnitude ratio of 4610 with $\Omega_{astro} = 3.55 \times 10^{-9}$ and a reference frequency of 25 Hz. 
Here we have a fitting uncertainty of $50\%$, which is the limit for making a measurement.
In fact, we can consider making a measurement of the cosmological background if the uncertainty is less than $50\%$; note the dashed line in Fig.~\ref{fig:Fisher2AET}. This example corresponds to a cosmological background of $\Omega_{Cosmo} = 7.7 \times 10^{-13}$  
In Fig.~\ref{fig:Fisher2AET} the same study is presented with four values for the astrophysical background: $\Omega_{astro} = 3.55 \times 10^{-8}$, $3.55 \times 10^{-9}$, $1.8 \times 10^{-9}$ and $3.55 \times 10^{-10}$. The same ratio produces similar results for different inputs of astrophysical amplitude. We obtain respectively the limits to contraining the cosmological background: $\Omega_{Cosmo} = 7.8 \times 10^{-12}$, $7.8 \times 10^{-13}$, $3.6 \times 10^{-13}$ and $7.6 \times 10^{-14}$. The value of these A-MCMC results are given in the Table~\ref{table:resultA+E+T}. Figs.~\ref{fig:Corner6paramAET} and \ref{fig:Corner6param2AET} present respective  examples of corner plots and posterior distributions for a run of a 6 parameter A-MCMC with $\Omega_{GW,Astro} = 3.55 \times 10^{-8}$ and  $\Omega_{GW,Cosmo} = 1 \times 10^{-10}$, $\Omega_{GW,Astro} = 3.55 \times 10^{-9}$ and $\Omega_{GW,Cosmo} = 5 \times 10^{-12}$. 

\section{Conclusion}
\label{sec:conclusion}
In this paper we present the potential for separating the spectral components of the two SGWBs with an adaptive MCMC method. We also implement a Fisher information study, predicting the measurement uncertainty from the A-MCMC analysis. The two independent studies produce consistent results. 
We obtained an uncertainty around 1 for the low level ($\Omega_0 = 1 \times 10^{-12}$) and around 0.03 for the high level ($\Omega_0 = 1 \times 10^{-8}$).
For example, with an astrophysical background of $\Omega_{GW,Astro} = 3.55 \times 10^{-9} \left(\frac{f}{25 \ \text{Hz}}\right)^{2/3}$ a cosmological background at $\Omega_{GW,Cosmo} = 7.6 \times 10^{-13}$ can be detected. This corresponds to an uncertainty $\frac{\Delta \Omega_0 }{\Omega_0}$ of $0.5$ (dashed line in the Fig.~\ref{fig:Fisher2AET}). The study presented in Sec.~\ref{sc:fisher+MCMC} displays the possibility to fit the parametric components of the SGWB. 

In the Sec.~\ref{sc:cov} we discussed and demonstrated the possibility to analyze the 'noise' channel (the $T$ channel) to fit the noise parameters of the LISA noise budget. The advantage of this method is to increase the efficiency of the parameter estimates and utilize the total frequency domain $[1 \times 10^{-5} \ \text{Hz}, 1 \ \text{Hz}]$. We also apply a Fisher information study to the LISA noise. According to the Fig.~\ref{fig:Fisher2AET} we show the possibility to separate the two SGWBs with a spectral separation with a factor of $4610$ (for $f_{ref} = 25 Hz$). Using a realistic range for the predicted magnitude of the astrophysically produced SGWB the methods demonstrated in this paper show that it is possible for LISA to also a observe a cosmologically produced SGWB in the range of $\Omega_{GW,Cosmo} \approx 1 \times 10^{-12}$ to $1 \times 10^{-13}$. 

We note some limitations in this study and give some expectations for future work. In this paper we assume no difference in the noise levels on each spacecraft. According to~\cite{Adams_2014} it is possible to include such a noise variation for each spacecraft. We could also include small modifications of the transfer functions $R_I$, and allow for some modification of the the spectral slopes of the noise components. We can have a varying slope but with a narrow Gaussian prior centered on the theoretical value. It will be important to address more detailed models of both the LISA noise and the astrophysical and cosmological contributions to the stochastic background.   

\section*{Acknowledgements}
\label{sec:awk}
The GB, NC and NJC thank the Centre national d'études spatiales (CNES) for support for this research. 
NJC appreciates the support of the NASA LISA Preparatory Science grant 80NSSC19K0320. 
RM acknowledges support by the James Cook Fellowship from Government funding, administered by the Royal Society Te Ap\={a}rangi and
DFG Grant KI 1443/3-2. 

\appendix
\section{Signal Separation Literature Summary}
\label{sec:appendix}
In this Appendix we present in tabular form a list of various studies that have been conducted in order to separate different SGWBs and detector noise sources. Much has been published on this subject.

\subsection{SGWB studies for LIGO/Virgo}
Table~\ref{table:introLIGO} presents a summary of the literature addressing SGWB signal separation for LIGO and Virgo.

\begin{table*}[!b]
\begin{ruledtabular}
\begin{tabularx}{\linewidth}{p{3.cm}|p{3.cm}|p{3.cm}|p{3.cm}|p{3.cm}}
 Reference & Goal & Method & Perfomance & Limitations~and applications \\ \hline \hline
Z.-C. Chen \textit{et. al.} \cite{2019ApJ...871...97C} & Astrophysical SGWB from binary black holes and binary neutrons stars & Estimation of the SGWB from LIGO/Virgo observations; local merger rate $\mathcal{R}$ & $\Omega_{2/3}=4.4^{+6.3}_{-3.0} \times 10^{-12} $, $f_{ref}=3$ mHz) & The error on the local merger rate is important \\ \hline
B. Abbott \textit{et. al.} \cite{PhysRevLett.116.131102} & Astrophysical SGWB from binary black holes and binary neutrons stars & Estimation of the SGWB from LIGO/Virgo observations with the local merger rate $\mathcal{R}$ estimation from GW150914 & $\Omega_{2/3}=1.1^{+2.7}_{-0.9} \times 10^{-12} $ $f_{ref}=25$ Hz & The error on the local merge rate is important \\ \hline

B. Abbott \textit{et. al.} \cite{LIGOScientific:2019vic,Abbott:2021xxi}& 3 backgrounds considered, power laws $\alpha=0, \frac{2}{3}, 3$ &Results  from  cross-correlation analysis with Advanced LIGO O3 combined O1 and O1 results&    $\Omega_{0} < 5.8\times 10^{-9}$; $\Omega_{2/3} < 3.4\times 10^{-9}$, $f_{ref}=25$ Hz & No correlated noise due to the magnetic Schumann resonances \\ \hline
 
A. Parida \textit{et. al.} \cite{2016JCAP...04..024P}&  Separate different isotropic SGWBs for LIGO &
Component separation of power laws avoiding use of MCMC methods& Simulation demonstration for Advanced LIGO target sensitivity: 
    $\Omega_{0} = (1 \pm 0.676)\times 10^{-8} $;
    $\Omega_{2/3} = (1 \pm 1.719)\times 10^{-8}$;  
     $\Omega_{3} = (1 \pm 3.284)\times 10^{-8}$; $f_{ref}=100$ Hz.
& Requires a negligible amount of computation and be simple to apply to real data \\ \hline

C. Ungarelli and A. Vecchio \cite{Ungarelli_2004}&Fit broken power-law SGWB with data from Earth-based detectors& Filters based on broken power-law spectra & Achieved fitting factor greater than 97\% &Small number of filters needed to measure SGWB in the first generation laser interferometers\\\hline

R. Smith and E. Thrane \cite{Smith:2017vfk} & To detect astrophysical SGWB with LIGO/Virgo & Bayesian parameter estimation to detect unresolved binary black hole background & Less data needed to observe background, as opposed to traditional correlation based search&Gives a unified method for a search for resolvable signals and a SGWB of unresolvable signals\\ \hline

S. Biscoveanu \textit{et. al.} \cite{Biscoveanu:2020gds}& To detect a primordial SGWB in the presence of unresoved binary black holes in LIGO/Virgo band & Use method of \cite{Smith:2017vfk}; individual short time segments analyzed & Measurement of a simulated power law: $ \log \Omega_{\alpha} = -5.96^{+0.08}_{-0.16}$, $ \alpha = 0.49^{+1.14}_{-0.49}$ 
& Limitations from the precision of the compact binary signal waveforms, and non-Gaussian noise \\ \hline

E. Thrane \textit{et. al.} \cite{Thrane:2013npa} &SGWB measurement in the context of correlated magnetic noise in LIGO/Virgo band. &Correlated noise between detectors creates a systematic error in cross correlation study &Measurement of the correlated noise from the Schumann resonances & Possibility to use Wiener Filter to subtract the correlation. \\\hline

P. M. Meyers \textit{et. al.} \cite{Meyers:2020qrb} &LIGO/Virgo SGWB measurement in the context of correlated magnetic noise &Parameter estimation of the correlated magnetic  noise and SGWB & Demonstration with  $\Omega_{2/3} \simeq3 \times10^{-9}$, $f_{ref}=25$ Hz and realistic magnetic coupling in LIGO/Virgo &  An alternative to Wiener filtering \\
\end{tabularx}
\caption{Methods to measure and to separate SGWBs for LIGO/Virgo.}
\label{table:introLIGO}
\end{ruledtabular}
\end{table*}

\subsection{SGWB studies in LISA band}
Table~\ref{table:introLISA} presents a summary of the literature addressing SGWB signal separation for LISA.

\begin{table*}[!b]
\begin{ruledtabular}
\begin{tabularx}{\linewidth}{p{3.cm}|p{3.cm}|p{3.cm}|p{3.cm}|p{3.cm}}
 Reference & Goal & Method & Perfomance & Limitations~and applications \\ \hline \hline
N. J. Cornish and S. L. Larson \cite{Cornish_2001} & Observe cosmic SGWB with astrophysical foregrounds & Strategies for individual,  or two LISA interferometers using cross-correlation & LISA could detect a cosmic SGWB at the level of $\Omega_{GW}(f)h_0^2 > 7 \times 10^{-12} $ & The LISA sensibility is derived for LISA arm-length of $L = 5 \times 10^9$ \\ \hline

 M. Pieroni and E. Barauss \cite{Pieroni_2020}& Extraction of the cosmological SGWB and astrophysical foreground with LISA noise& Principal component analysis to model and extract SGWBs& LISA can measure a cosmological SGWB of
$\Omega_{0}=6 \times 10^{-13}$ with $SNR = 31 $ 
& A robust technique that can be extended to different detectors  \\ \hline

C. Caprini \textit{et. al.} \cite{Caprini:2019pxz}&Observe SGWBs with LISA &Reconstruction of SGWB as a function of frequency for simple and broken power-laws  &Detects a  power law of $\Omega_{2/3}=5.4 \times 10^{-12}$, $f_{ref}=0.001$ Hz, with SNR=601. 
&Signal and noise are assumed to be stationary for all times.
\\\hline

R. Flauger \textit{et. al.} \cite{1818908} &Observe SGWBs with LISA, building on the work of \cite{Caprini:2019pxz} &Reconstruction of the spectral shape of a SGWB with the LISA $A, E, T$ channels &
Improvement of $\sqrt{2}$ over the method of \cite{Caprini:2019pxz}
&Will be expanded to account for unequal arm-lengths for LISA constellation.  \\\hline

N. Karnesis \textit{et. al.} \cite{10.1088/1361-6382/abb637}& Fast methodology to assess LISA detectability of a stationary, Gaussian, and isotropic SGWB &Testing the \textit{Radler} simulated dataset from the LISA Data Challenge & Successful demonstration for $\Omega_{2/3}(f) = 3.6\times 10^{-9}\left(\frac{f}{25 \ \text{Hz}}\right)^{2/3}$ &Analysis done with simple LISA noise model \\
\end{tabularx}
\caption{Methods to measure and to separate SGWBs for LISA }
\label{table:introLISA}
\end{ruledtabular}
\end{table*}

\subsection{SGWB studies for the future third generation detectors}
Table~\ref{table:intro3gene} presents a summary of the literature addressing SGWB signal separation for third generation gravitational-wave detectors.

\begin{table*}[!b]
\begin{ruledtabular}
\begin{tabularx}{\linewidth}{p{3.cm}|p{3.cm}|p{3.cm}|p{3.cm}|p{3.cm}}
 Reference & Goal & Method & Perfomance & Limitation~and application \\ \hline \hline

T. Regimbau \textit{et. al.}~\cite{PhysRevLett.118.151105}&Observing a primordial SGWB below the compact binary produced background &The data will be cleaned of the direct observations of binaries by the third generation detectors &Possible limit of $\Omega_{GW} \simeq 10^{-13}$ after 5 years of observation with third generation detectors~\cite{Punturo_2010,Reitze:2019iox}&Potential limitation to sensitivity comes from other astrophysical gravitational-wave emission. \\\hline

A. Sharma and J. Harms \cite{PhysRevD.102.063009}&Cosmological SGWB with third-generation detectors in the presence of an astrophysical foreground &  Matched filtering and residual study for the astrophysical foreground and cross-correlation for cosmological SGWB & Cosmological SGWB (flat) $\Omega_{GW} = 2 \times 10^{-12} $ observed with $SNR\approx 5.2$ after 1.3 years &Limitation for cosmological SGWB is instrumental noise and un-removed astrophysical sources\\ \hline

K. Martinovic \textit{et. al.}~\cite{Martinovic:2020}& Astrophysical (compact binary coalescence) and cosmological SGWB (cosmic strings and first order phase transitions) & Bayesian parameter estimation for simultaneous estimation of astrophysical and cosmological SGWB with third generation detectors & Possible limit at 25 Hz of  $\Omega_{GW} = 2.2 \times 10^{-13} $ (broken power-law model for primordial SGWB) and $\Omega_{GW} = 4.5. \times 10^{-13}$ for cosmic strings& Methods will be applicable for LISA\\

\end{tabularx}
\caption{Methods to measure and to separate SGWBs for the third generation detectors.}
\label{table:intro3gene}
\end{ruledtabular}
\end{table*}
\nocite{*}

\bibliography{biblio}
\bibliographystyle{abbrv}
\end{document}